\documentclass[submission]{SciPost}

\usepackage{cite}
\usepackage[pdftex]{graphicx}

\newcommand{\be}{\begin{equation}}
\newcommand{\ee}{\end{equation}}
\newcommand{\bea}{\begin{eqnarray}}
\newcommand{\eea}{\end{eqnarray}}

\begin{document}

\begin{center}{\Large \textbf{Quantum quenches to the attractive one-dimensional Bose gas: exact results}}\end{center}

\begin{center}
L. Piroli\textsuperscript{1*}, 
P. Calabrese\textsuperscript{1}, 
F.H.L. Essler\textsuperscript{2}
\end{center}

\begin{center}
{\bf 1} SISSA and INFN, via Bonomea 265, 34136 Trieste, Italy.
\\
{\bf 2} The Rudolf Peierls Centre for Theoretical Physics,
    Oxford University, Oxford, OX1 3NP, United Kingdom.
\\
* lpiroli@sissa.it
\bigskip
\end{center}


\section*{Abstract}
{\bf 
We study quantum quenches to the one-dimensional Bose gas with
attractive interactions in the case when the initial state is an ideal
one-dimensional Bose condensate. We focus on properties of the
stationary state reached at late times after the quench. This displays
a finite density of multi-particle bound states, whose rapidity
distribution is determined exactly by means of the quench action
method. We discuss the relevance of the multi-particle bound states for
the physical properties of the system, computing in particular the
stationary value of the local pair correlation function $g_2$. 
}

\section{Introduction}

Strongly correlated many-body quantum systems are often outside the
range of applicability of standard perturbative methods. While being
at the root of many interesting and sometimes surprising physical
effects, this results in huge computational challenges, which are most
prominent in the study of the non-equilibrium dynamics of many-body
quantum systems. 
This active field of research has attracted increasing attention over
the last decade, also due to the enormous experimental advances in
cold atomic physics \cite{bloch, polkovnikov, cazalilla}. Indeed,
highly isolated many-body quantum systems can now be realised in cold
atomic laboratories, where the high experimental control allows to
directly probe their unitary time evolution  \cite{kinoshita-06,
  cheneau, gring,trotzky, fukuhara, langen-13, agarwal, geiger,
  langen-15,kauf}.  

A simple paradigm to study the non-equilibrium dynamics of closed
many-body quantum systems is that of a quantum quench \cite{cc-05}: a
system is prepared in an initial state (usually the ground state of
some Hamiltonian $H_0$) and it is subsequently time evolved with a
local Hamiltonian $H$. In the past years, as a result of a huge
theoretical effort (see the reviews  
 \cite{polkovnikov,efg-14,dkpr-15,ge-15,a-16,ef-16,c-16,cc-16} and
 references therein), a robust picture has emerged: at
 long times after the quench, and in the thermodynamic limit,
 expectation values of {\it local} observables become stationary. For 
 a generic system, these stationary values are those of a thermal
 Gibbs ensemble with the effective temperature being fixed by the
energy density in the initial state \cite{rdo-08}.  

A different behaviour is observed for integrable quantum systems,
where an infinite set of local conserved charges constrains the
non-equilibrium dynamics. In this case, long times after the quench, 
local properties of the systems are captured by a generalised
Gibbs ensemble (GGE) \cite{rdyo-07}, which is a natural extension of
the Gibbs density matrix taking into account a complete set of local
or quasi-local conserved charges. 

The initial focus was on the role played by (ultra-)local conservation
laws in integrable quantum spin
chains\cite{cef-11,ck-12,fe-13,bp-13a,fe-13b,kscc-13,fcec-14}, while
more recent works have clarified the role by sets of novel,
quasi-local charges\cite{prosen-11,pi-13,prosen-14,
  ppsa-14,fagotti,imp-15,doyon-15,zmp-16,pv-16,fagotti-16, impz-16}. 
It has been shown recently that they have to be taken into account in
the GGE construction in order to obtain a correct description of
local properties of the steady state
\cite{idwc-15,iqdb-15}. Quasi-local conservation laws and their
relevance for the GGE have also recently been discussed in the
framework of integrable quantum field theories
\cite{emp-15,cardy-15}. These works have demonstrated that the problem
of determining a complete set of local or quasi-local conserved
charges is generally non-trivial.  

A different approach to calculating expectation values of local
correlators in the stationary state was introduced in
Ref.~\cite{ce-13}. It is the so called quench action method (QAM),
a.k.a. representative eigenstate approach and it does not rely on the
knowledge of the conserved charges of the system. Within this method,
the local properties at large times are effectively described by a
single eigenstate of the post-quench Hamiltonian. The QAM has now been
applied to a variety of quantum quenches, from one dimensional Bose
gases \cite{dwbc-14, dc-14, dpc-15,vwed-15, bucciantini-15, pce-16} to
spin chains \cite{wdbf-14, bwfd-14, budapest,buda2,dmv-15} and integrable
quantum field theories \cite{bse-14,bpc-16}, see Ref. \cite{c-16} for
a recent review. 

One of the most interesting aspects of non-equilibrium dynamics in
integrable systems is the possibility of realising non-thermal, stable
states of matter by following the unitary time evolution after a
quantum quench. Indeed, the steady state often exhibits properties
that are qualitatively different from those of thermal states of
the post-quench Hamiltonian. The QAM provides a powerful tool to
theoretically investigate these properties in experimentally relevant
settings. 

In this paper we study the quantum quench from an ideal Bose
condensate to the Lieb-Liniger model with arbitrary attractive
interactions. A brief account of our results was previously given in
Ref.~\cite{pce-16}. The interest in this quench lies in its
experimental feasibility as well as in the intriguing features of the
stationary state, which features finite densities of multi-particle
bound states. Our treatment, based on the quench action method, allows
us to study their dependence on the final interaction strength and
discuss their relevance for the physical properties of the system. In
particular, as a meaningful example, we consider the local pair
correlation function $g_2$, which we compute exactly.  

The structure of the stationary state is very different from 
the super Tonks-Girardeau gas, which is obtained by quenching the
one-dimensional Bose gas from infinitely repulsive to infinitely
attractive interaction \cite{abcg-05, bbgo-05,hgmd-09, mf-10, kmt-11,
  pdc-13, th-15}. The super Tonks-Girardeau gas features no bound
states, even though it is more strongly correlated than the infinitely
repulsive Tonks-Girardeau gas, as has been observed experimentally
\cite{hgmd-09}. As we argued in \cite{pce-16}, the physical properties
of the post-quench stationary state reached in our quench protocol
could be probed in ultracold atoms experiments, and the multi-particle
bound states observed by the presence of different``light-cones''
in the spreading of local correlations following a local quantum quench
\cite{gree-12}.

In this work we present a detailed derivation of the results
previously announced in Ref.~\cite{pce-16}. The remainder of this 
manuscript is organised as follows. In section~\ref{model} we
introduce the Lieb-Liniger model and the quench protocol that we
consider. The quench action method is reviewed in section~\ref{QAM},
and its application to our quench problem is detailed. In
section~\ref{stationary_eq} the equations describing the post-quench
stationary state are derived. Their solution is then obtained in 
section~\ref{exact_solution}, and a discussion of its properties is
presented. In section~\ref{phys_prop} we address the calculation of
expectation values of certain local operators on the post-quench
stationary state, and we explicitly compute the local pair correlation
function $g_2$. Finally, our conclusions are presented in
section~\ref{conclusions}. For the sake of clarity, some technical
aspects of our work are consigned to several appendices.

\section{The Lieb-Liniger model}
\label{model}
\subsection{The Hamiltonian and the eigenstates}
\label{eigenstates}
We consider the Lieb-Liniger model \cite{lieb}, consisting of $N$
interacting bosons on a one-dimensional ring of circumference $L$. The
Hamiltonian reads  
\begin{equation}
H^N_{LL}=-\frac{\hbar^2}{2m}\sum_{j=1}^{N}\frac{\partial^2}{\partial x_{j}^2}+2c\sum_{j<k}\delta(x_j-x_k),
\label{hamiltonian}
\end{equation}
where $m$ is the mass of the bosons, and $c=-\hbar^2/ma_{1D}$ is the
interaction strength. Here $a_{1D}$ is the 1D effective scattering
length \cite{olshanii-98} which can be tuned via Feshbach resonances
\cite{iasm-98}. In the following we fix $\hbar=2m=1$. The second
quantized form of the Hamiltonian is
\begin{equation}
H_{LL}=\int_{0}^{L}\mathrm{d}x \Big\{\partial_x\Psi^{\dagger}(x)\partial_x\Psi(x)+c\Psi^{\dagger}(x)\Psi^{\dagger}(x)\Psi(x)\Psi(x)\Big\},
\label{hamiltonian_2}
\end{equation}
where $\Psi^{\dagger}$, $\Psi$ are complex bosonic fields satisfying
$[\Psi(x),\Psi^{\dagger}(y)]=\delta(x-y)$. 

The Hamiltonian (\ref{hamiltonian}) can be exactly diagonalised for
all values of $c$ using the Bethe ansatz method
\cite{lieb,korepin}. Throughout this work we will consider the
attractive regime $c<0$ and use notations
$\overline{c}=-c>0$. We furthermore define a dimensionless coupling
constant by
\begin{equation}
\gamma=\frac{\overline{c}}{D}\ ,\quad D=\frac{N}{L}.
\end{equation}

A general $N$-particle energy eigenstate is parametrized by a set of
$N$ complex rapidities $\{\lambda_j\}_{j=1}^{N}$, satisfying the
following system of Bethe equations
\begin{equation}
e^{-i\lambda_jL}=\prod_{k\neq j\atop k=1}^{N}\frac{\lambda_k-\lambda_j-i\overline{c}}{\lambda_k-\lambda_j+i\overline{c}}\ ,\quad j=1,\ldots, N\ .
\label{bethe_eq}
\end{equation}
The wave function of the eigenstate corresponding to the set of rapidities
$\{\lambda_j\}_{j=1}^{N}$ is then 
\begin{equation}
\psi_N(x_1,\ldots,x_N|\{\lambda_j\}_{j=1}^N)=\frac{1}{\sqrt{N}}\sum_{P\in \mathcal{S}_N}e^{i\sum_{j}x_j\lambda_{P_j}} \prod_{j>k}\frac{\lambda_{P_j}-\lambda_{P_k}+i\overline{c}\mathrm{sgn}(x_j-x_k)}{\lambda_{P_j}-\lambda_{P_k}},
\end{equation}
where the sum is over all the permutations of the rapidities. Eqns
(\ref{bethe_eq}) can be rewritten in logarithmic form as 
\begin{equation}
\lambda_jL-2\sum_{k=1}^{N}\arctan\left(\frac{\lambda_j-\lambda_k}{\overline{c}}\right)=2\pi I_j\ ,\quad j=1,\ldots, N\ ,
\label{bethe_log}
\end{equation}
where the quantum numbers $\{I_j\}_{j=1}^{N}$ are integer (half-odd
integer) for $N$ odd (even).

In the attractive regime the solutions of (\ref{bethe_log}) organize themselves into mutually disjoint patterns in the complex rapidity plane called ``strings'' \cite{takahashi, cc-07}. For a given $N$ particle state, we indicate with $\mathcal{N}_s$ the total number of strings and with $N_j$ the number of $j$-strings, i.e. the strings containing $j$ particles ($1\leq j\leq N$) so that
\begin{equation}
N=\sum_{j}jN_j,\qquad \mathcal{N}_s=\sum_{j}N_j.
\end{equation} 
The rapidities within a single $j$-string are parametrized as\cite{mg-64}
\begin{equation}
\lambda^{j,a}_{\alpha}=\lambda_{\alpha}^{j}+\frac{i\overline{c}}{2}(j+1-2a)+i\delta^{j,a}_{\alpha} ,\quad a=1,\ldots, j ,
\label{structure}
\end{equation}
where $a$ labels the individual rapidities within the $j$-string, while
$\alpha$ labels different strings of length $j$. Here
$\lambda_{\alpha}^j$ is a real number called the string centre. The 
structure (\ref{structure}) is common to many integrable systems and
within the so called string hypothesis \cite{takahashi, thacker} the
deviations from a perfect string $\delta^{j,a}_{\alpha}$ are assumed
to be exponentially vanishing with the system size $L$ (see
Refs.~\cite{sakmann, sykes} for a numerical study of such deviations
in the Lieb-Liniger model). A $j$-string can be seen to correspond to
a bound state of $j$ bosons: indeed, one can show that the Bethe
ansatz wave function decays exponentially with respect to the distance
between any two particles in the bound state and the $j$ bosons can be
thought as clustered together. 

Even though some cases are known where states violating the string
hypothesis are present \cite{vladimirov, essler-92, ilakovac, fujita,
  hagemans}, it is widely believed that their contribution to
physically relevant quantities is vanishing in the thermodynamic
limit. We will then always assume the deviations
$\delta^{j,a}_{\alpha}$ to be exponentially small in $L$ and neglect
them except when explicitly said otherwise. 
 
From (\ref{bethe_log}), (\ref{structure}) a system of equations for
the string centres $\lambda^j_{\alpha}$ is obtained \cite{cc-07} 
\begin{equation}
j\lambda_{\alpha}^{j}L-\sum_{(k,\beta)}\Phi_{jk}(\lambda^{j}_{\alpha}-\lambda_{\beta}^{k})=2\pi I^{j}_{\alpha}\ ,
\label{BGT}
\end{equation}
where
\bea
\Phi_{jk}(\lambda)&=&(1-\delta_{jk})\phi_{|j-k|}(\lambda)+2\phi_{|j-k|+2}(\lambda)+\ldots+2\phi_{j+k-2}(\lambda)+\phi_{j+k}(\lambda)\ ,
\\
\phi_j(\lambda)&=&2\arctan\left(\frac{2\lambda}{j\overline{c}}\right)\ ,
\eea
and where $I^j_{\alpha}$ are integer (half-odd integer) for $N$ odd (even).
Eqns (\ref{BGT}) are called Bethe-Takahashi equations
\cite{takahashi,gaudin}. The momentum and the energy of a general
eigenstate are then given by 
\be
K=\sum_{(j,\alpha)} j \lambda^j_{\alpha}\ ,\qquad E = \sum_{(j,\alpha)} j (\lambda^j_{\alpha})^2 - \frac{\bar c^2}{12} j(j^2 - 1).
\label{momentum_energy}
\ee

\subsection{The thermodynamic limit}
In the repulsive case the thermodynamic limit 
\begin{equation}
N,L\to \infty\ ,\quad
D=\frac{N}{L}\ {\rm fixed},
\end{equation}
was first considered in Ref.~\cite{yy-69}, and it is well studied in
the literature. In the attractive case, the absolute value of the
ground state energy in not extensive, but instead grows as $N^3$ 
\cite{mg-64,cd-75}. While ground state correlation functions can be
studied in the zero density limit, namely $N$ fixed, $L\to \infty$
\cite{cc-07}, it was argued that the model does not have a proper
thermodynamic limit in thermal equilibrium
\cite{cd-75,takahashi}. Crucially, in the quench protocol we are 
considering, the energy is fixed by the initial state and the limit of
an infinite number of particles at fixed density presents no problem. 

As the systems size $L$ grows, the centres of the strings associated
with an energy eigenstate become a dense set on the real line and in the
thermodynamic limit are described by smooth distribution function. 
In complete analogy with the standard finite-temperature formalism
\cite{takahashi} we introduce the distribution function
$\{\rho_n(\lambda)\}_{n=1}^{\infty}$ describing the centres of $n$
strings, and the distribution function of holes
$\{\rho^h_n(\lambda)\}_{n=1}^{\infty}$. We recall that
$\rho^h_n(\lambda)$ describes the distribution of unoccupied states for 
the centres of $n$-particle strings, and is analogous to the
distribution of holes in the case of ideal Fermi gases at finite
temperature. Following Takahashi~\cite{takahashi} we introduce
\bea 
\eta_{n}(\lambda)&=&\frac{\rho_{n}^{h}(\lambda)}{\rho_{n}(\lambda)},\label{eq:eta}\\
\rho^{t}_{n}(\lambda)&=&\rho_n(\lambda)+\rho_{n}^h(\lambda). \label{rho_tot}
\eea
In the thermodynamic limit the Bethe-Takahashi equations (\ref{BGT})
reduce to an infinite set of coupled, non-linear integral equations
\begin{equation}
\frac{n}{2\pi}-\sum_{m=1}^{\infty}\int_{-\infty}^{\infty}\mathrm{d}\lambda'a_{nm}(\lambda-\lambda')\rho_{m}(\lambda')=\rho_{n}(\lambda)(1+\eta_{n}(\lambda)).
\label{coupled}
\end{equation}
where
\bea
a_{nm}(\lambda)&=&(1-\delta_{nm})a_{|n-m|}(\lambda)+2a_{|n-m|+2}(\lambda)+\ldots+2a_{n+m-2}(\lambda)+a_{n+m}(\lambda)\ ,
\label{aa_function}\\
 a_{n}(\lambda)&=&\frac{1}{2\pi}\frac{\mathrm{d}}{\mathrm{d}\lambda}\phi_{n}(\lambda)=\frac{2}{\pi n \overline{c}}\frac{1}{1+\left(\frac{2\lambda}{n \overline{c}}\right)^2}\ .
\label{a_function}
\eea
In the thermodynamic limit the energy and momentum per volume are
given by
\bea
k[\{\rho_n\}]=\sum_{n=1}^{\infty}\int_{-\infty}^{\infty}{\rm d}\lambda\ \rho_n(\lambda)n\lambda,\qquad e[\{\rho_n\}]=\sum_{n=1}^{\infty}\int_{-\infty}^{\infty}{\rm d}\lambda\ \rho_n(\lambda)\varepsilon_n(\lambda),
\label{therm_momentum_energy}
\eea
where 
\be
\varepsilon_{n}(\lambda)=n \lambda^2 - \frac{\bar c^2}{12} n(n^2 - 1).
\label{eq:epsilon}
\ee
Finally, it is also useful to define the densities $D_n$ and energy densities
$e_n$ of particles forming $n$-strings
\be
D_n= n\int_{-\infty}^{\infty}{\rm d}\lambda\ \rho_n(\lambda),\qquad 
e_n=\int_{-\infty}^{\infty}{\rm d}\lambda\ \rho_n(\lambda)\varepsilon_n(\lambda).
\label{density_per_string}
\ee
The total density and energy per volume are then additive
\be
D=\sum_{n=1}^{\infty}D_n,\qquad e=\sum_{n=1}^{\infty}e_n.
\label{Dtot}
\ee

\subsection{The quench protocol}
We consider a quantum quench in which the system is initially prepared
in the BEC state, i.e. the ground state of (\ref{hamiltonian}) with
$c=0$, and the subsequent unitary time evolution is governed by the
Hamiltonian (\ref{hamiltonian}) with $c=-\overline{c}<0$. The same
initial state was considered for quenches to the repulsive Bose
gas in Refs~\cite{kscc-13, dwbc-14, dc-14, kcc-14, grd-10, zwkg-15},
while different initial conditions were considered in
Refs~\cite{m-13,mossel-c-12, ia-12, csc-13, ga-14,sc-14, mckc-14,
  fgkt-15, goldstein-15, bucciantini-15, cgfb-14,gfcb-16}. 

As we mentioned before, the energy after the quench is conserved and
is most easily computed in the initial state $|\psi(0)\rangle=|{\rm BEC}\rangle$ as 
\be
\langle {\rm BEC}|H_{LL}|{\rm BEC}\rangle= -\overline{c}\langle{\rm BEC}|\int_0^{L}{\rm d}x\ \Psi^{\dagger}(x)\Psi^{\dagger}(x)\Psi(x)\Psi(x)|{\rm BEC}\rangle.
\ee 
The expectation value on the r.h.s. can then be easily computed using
Wick's theorem. In the thermodynamic limit we have
\be
\frac{E}{L}=-\overline{c}D^2=-\gamma D^3.
\label{energy}
\ee

\section{The quench action method}
\label{QAM}
\subsection{General considerations}
Consider the post-quench time evolution of the expectation value of a general
operator ${O}$. For a generic system it can be written as 
\be
\langle \psi(t) | {O} | \psi(t) \rangle= \sum_{\mu, \nu} \langle
\psi(0)|\mu\rangle \langle \mu|{O}|\nu\rangle\langle \nu | \psi(0)\rangle
e^{i (E_\mu-E_\nu)t},
\label{ds}
\ee
where $\{|\mu\rangle\}$ denotes an orthonormal basis of eigenstates of
the post-quench Hamiltonian. In Ref.~\cite{ce-13} it was argued that in
integrable systems a major simplification occurs if one is interested
in the time evolution of the expectation values of {\it local}
operators $\mathcal{O}$ in the thermodynamic limit. In particular, the
double sum in the spectral representation (\ref{ds}) can be
replaced by a single sum over particle-hole excitations over
a \emph{representative eigenstate} $|\rho_{sp}\rangle$. In
particular, we have  
\be
{\rm lim}_{\rm th}\langle \psi(t) | \mathcal{O} | \psi(t) \rangle= \frac{1}{2}\sum_{{\rm {\bf e}}}\left(e^{-\delta s_{{\rm {\bf e}}} - i\delta \omega_{{\rm {\bf e}}} t}\langle \rho_{sp}|\mathcal{O}|\rho_{sp},{\rm {\bf e}}\rangle + e^{-\delta s^{\ast}_{{\rm {\bf e}}} + i\delta \omega_{{\rm {\bf e}}} t}\langle \rho_{sp},{\rm {\bf e}}|\mathcal{O}|\rho_{sp}\rangle \right),
\label{time_ev}
\ee
where we have indicated with ${\rm lim}_{\rm th}$ the thermodynamic
limit $N,L\to \infty$, keeping the density $D=N/L$ fixed. Here {\bf e}
denotes a generic excitation over the representative state
$|\rho_{sp}\rangle$. Finally we have
\be
\delta s_{{\rm {\bf e}}}=-\ln\frac{\langle \rho_{sp}, {\rm {\bf e}}|\psi(0)\rangle}{\langle \rho_{sp}|\psi(0)\rangle}, \qquad \delta\omega_{{\rm {\bf e}}}=\omega[\rho_{sp},{\rm {\bf e}}]-\omega[\rho_{sp}],
\ee
where $\omega[\rho_{sp}]$, $\omega[\rho_{sp},{\rm {\bf e}}]$ are the
energies of $|\rho_{sp}\rangle$ and $|\rho_{sp}, {\rm {\bf
    e}}\rangle$ respectively. The representative eigenstate (or 
``saddle-point state'') $|\rho_{sp}\rangle$ is described in the
thermodynamic limit by two sets of distribution functions
$\{\rho_{n}(\lambda)\}_{n}$, $\{\rho^{h}_{n}(\lambda)\}_{n}$. In
Ref.~\cite{ce-13} it was argued that these are selected by the
saddle-point condition 
\be
\frac{\partial S_{QA}[\rho]}{\partial \rho_n(\lambda)} \Big|_{\rho=\rho_{sp}}=0, \qquad n\geq 1,
\label{oTBA1}
\ee
where $S_{QA}[\rho]$ is the so-called quench action
\be 
S_{QA}[\rho]=2S[\rho]-S_{YY}[\rho].
\label{SQA}
\ee
Here $\rho$ is the set of distribution functions corresponding to
a general macro-state, $S[\rho]$ gives the thermodynamically leading
part of the logarithm of the overlap 
\be
S[\rho]=-{\rm lim}_{\rm th}{\rm Re}\ln\langle \psi(0)|\rho\rangle,
\label{therm_overlap}
\ee
and $S_{YY}$ is the Yang-Yang entropy. As we will see in section
 \ref{overlap_bec}, we will only have to consider parity-invariant
 Bethe states, namely eigenstates of the Hamiltonian
 (\ref{hamiltonian}) characterised by sets of rapidities satisfying
 $\{\lambda_j\}_{j=1}^{N}=\{-\lambda_j\}_{j=1}^{N}$. Restricting to
 the sector of the Hilbert space of parity invariant Bethe states, the
 Yang-Yang entropy reads 
\be
\frac{S_{YY}[\rho]}{L}= \frac{1}{2}\sum_{n=1}^\infty \int_{-\infty}^{\infty} d\lambda [\rho_n \ln (1+\eta_n)+ \rho_n^h \ln (1+\eta_n^{-1})].
\label{yang}
\ee
We note the global pre-factor $1/2$. From Eq. (\ref{time_ev}) it
follows that the saddle-point state $|\rho_{sp}\rangle$ can be seen
as the effective stationary state reached by the system at long
times. Indeed, if $\mathcal{O}$ is a local operator,
Eq. (\ref{time_ev}) gives 
\begin{equation}
\lim_{t\to\infty}{\rm lim}_{\rm th}\langle \psi(t) | \mathcal{O} | \psi(t) \rangle =\langle \rho_{sp}|\mathcal{O}|\rho_{sp}\rangle .
\end{equation}

\subsection{Overlaps with the BEC state}
\label{overlap_bec}
The main difficulty in applying the quench action method to a generic
quantum quench problems is the computation of the overlaps
$\langle\psi(0)| \rho\rangle$ between the initial state and eigenstates of the 
post-quench Hamiltonian. At present this problem has been solved only
in a small number of special cases \cite{ce-13,mossel-10, pozsgay-14,
  amsterdam_overlaps, brockmann, pc-14, hst-15,
  mazza-15,fz-16,cl-14}.

A conjecture for the overlaps between the BEC state and the Bethe
states in the Lieb-Liniger model first appeared in Ref.~\cite{dwbc-14}
and it was then rigorously proven, for arbitrary sign of the particle
interaction strength, in Ref.~\cite{brockmann}. As we have already
mentioned, the overlap is non-vanishing only for parity invariant
Bethe states, namely eigenstates characterised by sets of rapidities
satisfying $\{\lambda_j\}_{j=1}^{N}=\{-\lambda_j\}_{j=1}^{N}$
\cite{amsterdam_overlaps}. The formula reads 
\be
\langle \{\lambda_{j}\}_{j=1}^{N/2}\cup \{-\lambda_{j}\}_{j=1}^{N/2} |{\rm BEC}\rangle=\frac{\sqrt{(cL)^{-N}N!}}{\prod_{j=1}^{N/2}\frac{\lambda_j}{c}\sqrt{\frac{\lambda_j^2}{c^2}+\frac{1}{4}}}\frac{\mathrm{det}^{N/2}_{j,k,=1}G_{jk}^{Q}}{\sqrt{\mathrm{det}^{N}_{j,k,=1}G_{jk}}},
\label{overlap}
\ee
where
\be
G_{jk}=\delta_{jk}\left[L+\sum_{l=1}^{N}K(\lambda_j-\lambda_l)\right]-K(\lambda_j-\lambda_k),
\ee
\be
G^Q_{jk}=\delta_{jk}\left[L+\sum_{l=1}^{N/2}K^Q(\lambda_j,\lambda_l)\right]-K^Q(\lambda_j,\lambda_k),
\ee
\be
K^Q(\lambda,\mu)=K(\lambda-\mu)+K(\lambda+\mu),\qquad K(\lambda)=\frac{2c}{\lambda^2+c^2}.
\ee
The extensive part of the logarithm of the overlap (\ref{overlap}) was
computed in Ref.~\cite{dwbc-14} in the repulsive regime. A key
observation was that the ratio of the determinants is non-extensive, i.e. 
\be
{\rm lim}_{\rm th} \frac{\mathrm{det}^{N/2}_{j,k,=1}G_{jk}^{Q}}{\sqrt{\mathrm{det}^{N}_{j,k,=1}G_{jk}}}=\mathcal{O}(1).
\ee

In the attractive regime additional technical difficulties arise,
because matrix elements of the Gaudin-like matrices $G_{jk}$,
$G^{Q}_{jk}$ can exhibit singularities when the Bethe state contains
bound states \cite{cl-14}. This is analogous to the situation
encountered for a quench from the N\'{e}el state to the gapped XXZ
model \cite{wdbf-14, bwfd-14,budapest,buda2}. In particular, one can
see that the kernel $K(\mu-\nu)$ diverges as
$1/(\delta_{\alpha}^{n,a}-\delta_{\alpha}^{n,a+1})$ for two
``neighboring'' rapidities in the same string
$\mu=\lambda_{\alpha}^{n,a}$, $\nu=\lambda_{\alpha}^{n,a+1}$, or when 
rapidities from different strings approach one another in the
thermodynamic limit, $\mu\to\lambda+ic$. 

These kinds of singularities are present in the determinants of both
$G^{Q}_{jk}$ and $G_{jk}$. It was argued in Refs~\cite{wdbf-14,
  bwfd-14,cl-14} that they cancel one another in the expression for
the overlap. As was noted in Refs.~\cite{wdbf-14,bwfd-14,cl-14}, no
other singularities arise as long as one considers the overlap between
the BEC state and a Bethe state without zero-momentum $n$-strings,
(strings centred at $\lambda=0$). Concomitantly the ratio of the
determinants in (\ref{overlap}) is expected to give a sub-leading
contribution in the thermodynamic limit, and can be dropped. The
leading term in the logarithm of the overlaps can then be easily
computed along the lines of Refs.~\cite{wdbf-14, bwfd-14}
\be
\ln \langle\rho|{\rm BEC}\rangle=-\frac{LD}{2}\left(\ln\gamma+1\right)+\frac{L}{2}\sum_{m=1}^{\infty}\int_{0}^{\infty}d\lambda \rho_{n}(\lambda)\ln W_{n}(\lambda),
\label{leading}
\ee
where 
\be
W_n(\lambda)=\frac{1}{\frac{\lambda^2}{\overline{c}^2}\left(\frac{\lambda^2}{\overline{c}^2}+\frac{n^2}{4}\right)\prod_{j=1}^{n-1}\left(\frac{\lambda^2}{\overline{c}^2}+\frac{j^2}{4}\right)^2}.
\label{w_n}
\ee
In the case where zero-momentum $n$-strings are present, a more
careful analysis is required in order to extract the leading term of
the overlap (\ref{overlap}) \cite{cl-14,ac-15}. This is reported in Appendix~\ref{app_overlap}. The upshot of this analysis is that (\ref{leading})
gives the correct leading behaviour of the overlap even in the
presence of zero-momentum $n$-strings.

\section{Stationary state}
\label{stationary_eq}
\subsection{Saddle point equations}
As noted before, the stationary state is characterized by two sets
of distribution functions $\{\rho_n(\lambda)\}_n$, $\{\rho^h_n(\lambda)\}_n$,
which fulfil two infinite systems of coupled, non-linear integral
equations.   
The first of these is the thermodynamic version of the Bethe-Takahashi
equations (\ref{coupled}). The second set is derived from the
saddle-point condition of the quench action (\ref{oTBA1}), and the
resulting equations are sometimes called the overlap thermodynamic
Bethe ansatz equations (oTBA equations). Their derivation follows
Refs~\cite{wdbf-14,bwfd-14,budapest,buda2}. In order to fix the density
$D=N/L$ we add the following term to the quench action (\ref{SQA}) 
\be
-hL\left(\sum_{m=1}^{\infty}m\int_{-\infty}^{\infty}d\lambda\rho_{m}(\lambda)-D\right).
\label{density_condition}
\ee
As discussed in the previous section, $S[\rho]$ in (\ref{SQA}) can be
written as
\be
S[\rho]=\frac{LD}{2}\left(\ln\gamma+1\right)-\frac{L}{2}\sum_{m=1}^{\infty}\int_{0}^{\infty}d\lambda \rho_{n}(\lambda)\ln W_{n}(\lambda)\ ,
\label{s_term}
\ee
where $W_n(\lambda)$ is given in (\ref{w_n}). Using (\ref{s_term}),
(\ref{yang}), and (\ref{density_condition}) one can straightforwardly
extremize the quench action (\ref{SQA}) and arrive
at the following set of oTBA equations 
\begin{equation}
\ln\eta_{n}(\lambda)=-2hn-\ln W_{n}(\lambda)+\sum_{m=1}^{\infty}a_{nm}\ast \ln\left(1+\eta_{m}^{-1}\right)(\lambda),\qquad n\geq 1\ .
\label{coupled2}
\end{equation}
Here $a_{nm}$ are defined in (\ref{aa_function}), and we have used the notation
$f\ast g(\lambda)$ to indicate the convolution between two functions 
\be
f\ast g(\lambda)=\int_{-\infty}^{\infty}{\rm d}\mu \ f(\lambda-\mu)g(\mu).
\label{convolution}
\ee
Eqns (\ref{coupled2}) determine the functions $\eta_{n}(\lambda)$ and,
together with Eqns (\ref{coupled}) completely fix the distribution functions
$\{\rho_n(\lambda)\}_n$, $\{\rho^{h}_n(\lambda)\}_n$ characterising
the stationary state. 

\subsection{Tri-diagonal form of the oTBA equations}
Following standard manipulations of equilibrium TBA equations
\cite{takahashi}, we may re-cast the oTBA equations (\ref{coupled2})
in the form
\bea
\ln\eta_n(\lambda)=d(\lambda)+s\ast\left[\ln(1+\eta_{n-1})+\ln(1+\eta_{n+1})\right](\lambda)\ ,\qquad n\geq 1\ .
\label{finale_gtba}	
\eea
Here we have defined $\eta_0(\lambda)=0$ and
\bea
s(\lambda)=\frac{1}{2\overline{c}\cosh\left(\frac{\pi\lambda}{\overline{c}}\right)},\label{kernel}\\
d(\lambda)=\ln\left[\tanh^2\left(\frac{\pi\lambda}{2\overline{c}}\right)\right]. \label{a_driving}
\eea
The calculations leading to Eqns (\ref{finale_gtba}) are summarized in Appendix~\ref{app_tridiag}. 
The thermodynamic form of the Bethe-Takahashi equations
(\ref{coupled}) can be similarly rewritten. Since we do not make
explicit use of them in the following, we relegate their derivation to
Appendix~\ref{app_tridiag}.

\subsection{Asymptotic relations}
Eqns~(\ref{finale_gtba}) do not fix $\{\eta_{n}(\lambda)\}_n$ of
Eqns~(\ref{coupled2}), because they do not contain the chemical
potential $h$. In order to recover the (unique) solution of
Eqns~(\ref{coupled2}), it is then necessary  
to combine Eqns~(\ref{finale_gtba}) with a condition on the asymptotic
behaviour of $\eta_{n}(\lambda)$ for large $n$. In our case one can
derive from (\ref{coupled2}) the following relation, which holds
asymptotically for $n\to\infty$ 
\be
\ln\eta_{n+1}(\lambda)\simeq -2h+a_1\ast \ln\eta_n(\lambda)+\ln\left[\frac{\lambda}{\overline{c}}\left(\frac{\lambda^2}{\overline{c}^2}+\frac{1}{4}\right)\right].
\label{difference}
\ee
Here $a_1(\lambda)$ is given in (\ref{a_function}) (for $n=1$).  
The derivation of Eqn~(\ref{difference}) is reported in
Appendix~\ref{app_asymptotic}. The set of equations (\ref{finale_gtba}), with
the additional constraint given by Eqn~(\ref{difference}), is now
equivalent to Eqns~(\ref{coupled2}).

\section{Rapidity distribution functions for the stationary state}
\label{exact_solution}

\subsection{Numerical analysis}
\label{numerics}

Eqns (\ref{coupled}), (\ref{coupled2}) can be truncated
to obtain a finite system of integral equations, which are defined on
the real line $\lambda\in(-\infty,\infty)$. One can then numerically
solve this finite system either by introducing a cut-off for large
$\lambda$, or by mapping the equations onto a finite
interval. Following the latter approach, we define
\begin{equation}
\chi_{n}(\lambda)=\ln \left(\frac{\eta_{n}(\lambda)\tau^{2n}}{q_{n}(\lambda)}\right)\,,
\label{new_functions}
\end{equation}
where $q_{n}(\lambda)$ is given by
\be
q_n(\lambda)=\frac{1}{W_n(\lambda)}=\frac{\lambda^2}{\overline{c}^2}\left(\frac{\lambda^2}{\overline{c}^2}+\left(\frac{n}{2}\right)^2\right)\prod_{l=1}^{n-1}\left[\frac{\lambda^2}{\overline{c}^2}+\left(\frac{l}{2}\right)^2\right]^2\ .
\label{poly}
\ee
Finally, we have defined
\be
\tau=e^h,
\label{tau}
\ee
$h$ being the Lagrange multiplier appearing in (\ref{coupled2}).
The functions $\chi_n(\lambda)$ satisfy the following system of equations
\bea
\chi_{n}(\lambda)&=&\sum_{m=1}^{\infty}a_{nm}\ast \ln\left(1+\frac{\tau^{2m}}{q_m(\lambda)}e^{-\chi_m(\lambda)}\right)=\nonumber \\
&=&\sum_{m=1}^{\infty}\int_{0}^{+\infty}\mathrm{d}\ \mu\ (a_{nm}(\lambda-\mu)+a_{nm}(\lambda+\mu))\ln\left(1+\frac{\tau^{2m}}{q_m(\mu)}e^{-\chi_m(\mu)}\right),
\label{modified}
\eea
where $a_{nm}(\lambda)$ are defined in (\ref{aa_function}). We then
change variables
\be
\frac{\lambda(x)}{\overline{c}}=\frac{1-x}{1+x}\ , 
\label{map}
\ee
which maps the interval $(0,\infty)$ onto $(-1,1)$. Since the distributions $\chi_{n}(\lambda)$ are symmetric w.r.t. $0$, they can be described by functions with domain $(0,\infty)$. Using the map (\ref{map}) they become functions $\chi_{n}(x)$ with domain $(-1,1)$. The set of equations (\ref{modified}) becomes
\be
\chi_{n}(x)=2\sum_{m=1}^{\infty}\int_{-1}^{1}\ \mathrm{d}y \frac{1}{(1+y)^2}\mathcal{A}_{nm}(x,y)\ln\left(1+\frac{\tau^{2m}}{q_m(y)}e^{-\chi_m(y)}\right)\ ,
\label{compact1}
\ee
where
\be
\mathcal{A}_{nm}(x,y)=\overline{c}\left[a_{nm}\bigg(\lambda(x)-\lambda(y)\bigg)
+a_{nm}\bigg(\lambda(x)+\lambda(y)\bigg)\right]\,.
\label{new_function_2}
\ee
The thermodynamic Bethe-Takahashi equations (\ref{coupled}) can be
similarly recast in the form
\be
\Theta_{n}(x)=\frac{n}{2\pi}-2\sum_{m=1}^{\infty}\int_{-1}^{1}\frac{\mathrm{d}y}{(1+y)^2}\frac{\mathcal{A}_{nm}(x,y)}{1+\eta_{m}(y)}\Theta_{m}(y),
\label{compact2}
\ee
where $\Theta(x)=\rho^t_n\big(\lambda(x)\big)$, with $\lambda(x)$
defined in Eq. (\ref{map}). The infinite systems (\ref{compact1}) and
(\ref{compact2}), defined on the interval $(-1,1)$, can then be
truncated and solved numerically for the functions $\chi_n(x)$ and
$\Theta_n(x)$, for example using the Gaussian quadrature method. The
functions $\eta_n(\lambda)$ are recovered from (\ref{new_functions})
and (\ref{map}), while the particle and hole distributions
$\rho_n(\lambda)$, $\rho_n^h(\lambda)$ are obtained from the
knowledge of $\eta_n(\lambda)$ and $\rho_n^t(\lambda)$. 

\begin{figure}
\centering
\includegraphics[scale=0.93]{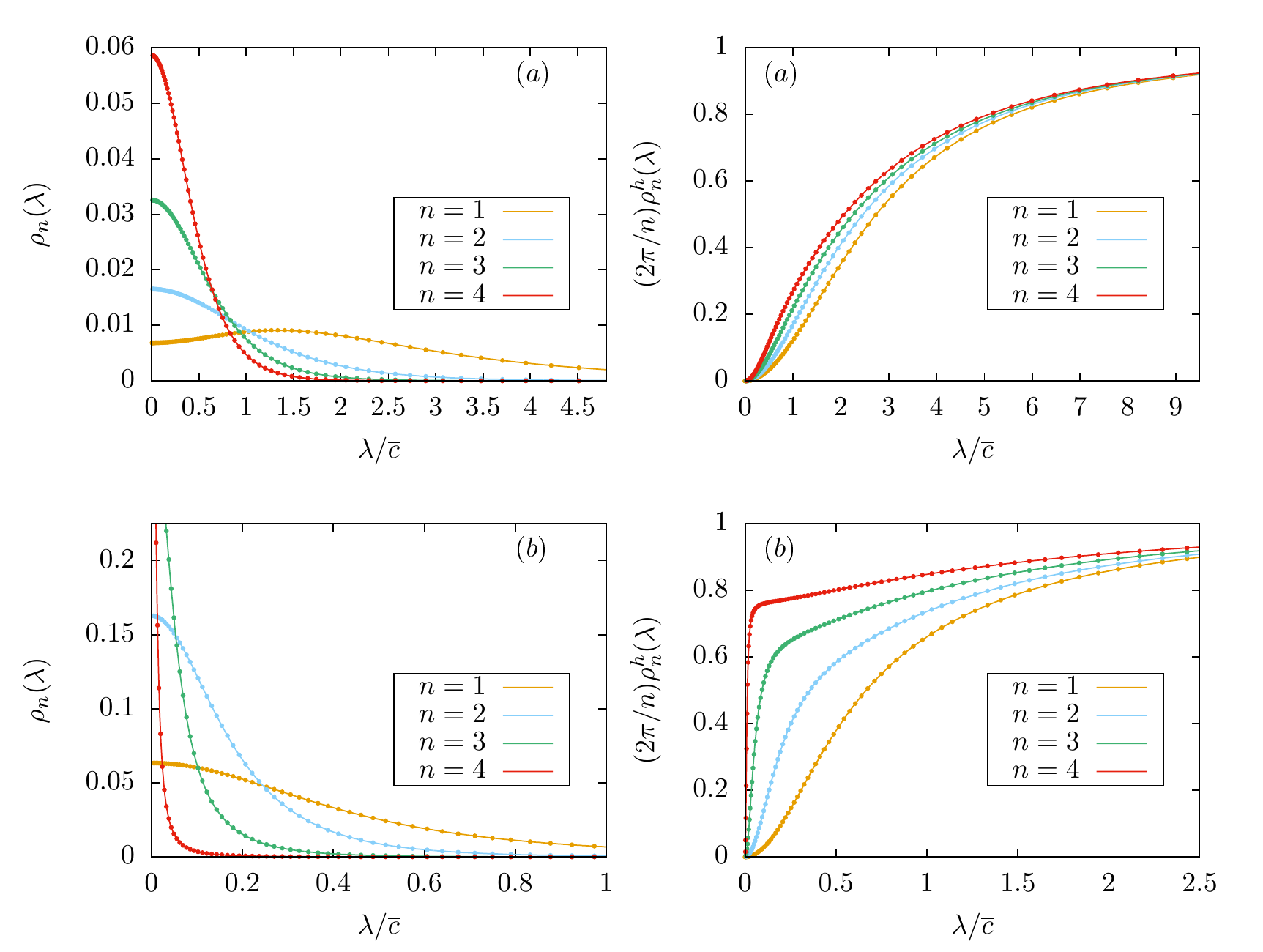}
\caption{Rapidity distribution functions $\rho_n(\lambda)$ and $(2\pi/
n)\rho_n^{h}(\lambda)$ for $n$-string solutions with $n\leq 4$. The
final value of the interaction is chosen as 
($a$) $\gamma=0.25$ and ($b$) $\gamma=2.5$. The dots correspond to
the numerical solution discussed in section~\ref{numerics}, while
solid lines correspond to the analytical solution presented in
section \ref{analytical}. The functions are shown for $\lambda>0$
(being symmetric with respect to $\lambda=0$) and have been rescaled
for presentational purposes. Note that the rescaling factors for the hole
distributions are determined by their asymptotic values, 
$\rho_n^h(\lambda)\to n/2\pi$ as $\lambda\to \infty$.} 
\label{distributions}
\end{figure}

As $\gamma$ decreases, we found that an increasing number of equations
has to be kept when truncating the infinite systems (\ref{compact1}),
(\ref{compact2}) in order to obtain an accurate numerical solution. As
we will see in section \ref{physical_discussions}, this is due to the
fact that, as $\gamma\to 0$, bound states with higher number of
particles are formed and the corresponding distribution functions
$\rho_n(\lambda)$, $\eta_n(\lambda)$ cannot be neglected in
(\ref{coupled}), (\ref{coupled2}). As an example, our numerical
solution for $\gamma=0.25$, and $\gamma=2.5$ is shown in
Fig.~\ref{distributions}, where we also provide a comparison with the
analytical solution discussed in section~\ref{analytical}.  

Two non-trivial checks for our numerical solution are available. 
The first is given by Eq.~(\ref{energy}), i.e. the solution has to
satisfy the sum rule
\be
\sum_{n=1}^{\infty}\int_{-\infty}^{\infty}{\rm d}\lambda\rho_n(\lambda)\varepsilon_n(\lambda)=-\gamma D^3,
\label{check1}
\ee
where $\varepsilon_n(\lambda)$ is defined in
Eqn (\ref{eq:epsilon}). The second non-trivial check was suggested in
Refs~\cite{budapest,buda2} (see also Ref.~\cite{bwfd-14}), and is
based on the observation that the action (\ref{SQA}) has to be equal to
zero when evaluated on the saddle point solution,
i.e. $S_{QA}[\rho_{sp}]=0$, or 
\be
2S[\rho_{sp}]=S_{YY}[\rho_{sp}],
\label{check2}
\ee
where $S[\rho]$ and $S_{YY}[\rho]$ are defined respectively in
(\ref{s_term}) and (\ref{yang}). Both (\ref{check1}) and
(\ref{check2}) are satisfied by our numerical solutions within a 
relative numerical error $\epsilon \lesssim 10^{-4}$ for all
numerically accessible values of $h$. As a final check we have
verified that our numerical solution satisfies, within numerical errors,
\begin{equation}
\gamma=\frac{1}{\tau}\ ,
\label{fact}
\end{equation}
where $\tau$ is defined in (\ref{tau}) and $\gamma=\bar{c}/D$ is computed from
the distribution functions using (\ref{Dtot}). Relation (\ref{fact})
is equivalent to that found in the repulsive case \cite{dwbc-14}.

\subsection{Perturbative expansion}
\label{perturbative_sec}
Following Ref.~\cite{dwbc-14} we now turn to a ``perturbative''
analysis of Eqns (\ref{coupled2}). This will provide us with another
non-trivial check on the validity of the analytical solution presented
in section \ref{analytical}.  
Defining $\varphi_n(\lambda)=1/\eta_n(\lambda)$ and using (\ref{tau}),
we can rewrite (\ref{coupled2}) in the form
\be
\ln \varphi_n(\lambda)=\ln(\tau^{2n})+\ln W_n(\lambda)-\sum_{m=1}^{\infty}a_{nm}\ast\ln(1+\varphi_m)(\lambda),
\label{perturbative}
\ee
where $W_n(\lambda)$ is given in (\ref{w_n}). We now expand the
functions $\varphi_n(\lambda)$ as power series in $\tau$
\be
\varphi_n(\lambda)=\sum_{k=0}^{\infty}\varphi^{(k)}_n(\lambda)\tau^k.
\ee
From (\ref{perturbative}) one readily sees that $\varphi_n(\lambda)=\mathcal{O}(\tau^{2n})$, i.e.
\bea
\varphi^{(k)}_n(\lambda)=0,\quad k=0,\ldots, 2n-1,\\
\varphi^{(2n)}_n(\lambda)=\frac{1}{\frac{\lambda^2}{\overline{c}^2}\left(\frac{\lambda^2}{\overline{c}^2}+\frac{n^2}{4}\right)\prod_{j=1}^{n-1}\left(\frac{\lambda^2}{\overline{c}^2}+\frac{j^2}{4}\right)^2}.
\label{coefficient}
\eea

Using (\ref{coefficient}) as a starting point we can now solve
Eqns (\ref{perturbative}) by iteration. The calculations are
straightforward but tedious, and are sketched in
Appendix~\ref{app_perturbative}. Using this method we have calculated
$\varphi_1(\lambda)$ up to fifth order in $\tau$. In terms of the
the dimensionless variable $x=\lambda/\overline{c}$ we have
\bea
\varphi_{1}(x)&=\frac{\tau^{2}}{x^2(x^2+\frac{1}{4})} 
\Bigg[1-\frac{4\tau}{x^{2}+1}+\frac{\tau^2(1+13x^2)}{(1+x^2)^2(x^2+\frac{1}{4})}-\frac{32(-1+5x^2)\tau^3}{(1+x^2)^3(1+4x^2)}\Bigg]+\mathcal{O}(\tau^{6}).
\label{fifth_order}
\eea

\subsection{Exact solution}
\label{analytical}
In this section we discuss how to solve equations (\ref{coupled}),
(\ref{coupled2}) analytically. We first observe that the distribution
functions $\rho_n(\lambda)$ can be obtained from the set
$\{\eta_n(\lambda)\}_n$ of functions fulfilling Eqns (\ref{coupled2}) as
\be
\rho_n(\lambda)=\frac{\tau}{4\pi}\frac{\partial_{\tau}\eta^{-1}_{n}(\lambda)}{1+\eta_n^{-1}(\lambda)},
\label{rho_n}
\ee 
where $\tau$ is given in (\ref{tau}). This relation is analogous to
the one found in the repulsive case in Ref.~\cite{dwbc-14}. To prove
(\ref{rho_n}) one takes the partial derivative $\partial_{\tau}$
of both sides of (\ref{perturbative}). Combining the resulting
equation with the thermodynamic version of the Bethe-Takahashi
equations (\ref{coupled}), and finally invoking the uniqueness of the
solution, we obtain (\ref{rho_n}). 

This leaves us with the task of solving (\ref{coupled2}). In what
follows we introduce the dimensionless parameter
$x=\lambda/\overline{c}$ and throughout this section, with a slight
abuse of notation, we will use the same notations for functions of
$\lambda$ and of $x$. Our starting point is the tri-diagonal form
(\ref{finale_gtba}) of the coupled integral equations (\ref{coupled2}).
Following Ref.~\cite{bwfd-14} we introduce the corresponding
$Y$-system \cite{suzuki, kp-92} 
\be
 y_{n}\left(x+\frac{i}{2}\right)y_{n}\left(x-\frac{i}{2}\right)=Y_{n-1}(x)Y_{n+1}(x), \qquad n\geq 1,
\label{y-system}
\ee
where we define $y_0(x)=0$ and
\be
Y_{n}(x)=1+y_{n}(x)\,.
\ee
Let us now assume that there exists a set of functions
$y_n(x)$ that satisfy the $Y$-system (\ref{y-system}), and as
functions of the complex variable $z$ have the following properties 
\begin{enumerate}
\item $y_{n}(z)\sim z^2$, as $z\to 0$, $\forall n\geq 1$; \label{prpty1}
\item $y_{n}(z)$ has no poles in $-1/2<\mathrm{Im}(z)<1/2$, $\forall n\geq 1$; \label{prpty2}
\item $y_{n}(z)$ has no zeroes in $-1/2<\mathrm{Im}(z)<1/2$ except for $z=0$, $\forall n\geq 1$. \label{prpty3}
\end{enumerate}
One can prove that the set of functions $y_n(x)$ with these properties
solve the tri-diagonal form  of the integral equations equations
(\ref{finale_gtba}) \cite{bwfd-14}. To see this, one has to first take 
the logarithmic derivative of both sides of (\ref{y-system}) and take the Fourier transform, integrating in $x\in(-\infty,\infty)$. Since the argument of the
functions in the l.h.s. is shifted by $\pm i/2$ in the imaginary
direction, one has to use complex analysis techniques to perform the
integral. In particular, under the assumptions (\ref{prpty1}),
(\ref{prpty2}), (\ref{prpty3}) the application of the residue theorem
precisely generates, after taking the inverse Fourier transform, the driving term $d(\lambda)$ in (\ref{finale_gtba}) \cite{bwfd-14}. 

We conjecture that the exact solution for $\eta_{1}(x)$ is given by
\begin{equation}
\eta_{1}(x)=\frac{x^2[1+4\tau+12\tau^2+(5+16\tau)x^2+4x^4]}{4\tau^2(1+x^2)}\,.
\label{eta_1}
\end{equation}
Our evidence supporting this conjecture is as follows:
\begin{enumerate}
\item{} We have verified using Mathematica that the functions
$\eta_n(x)$ generated by substituting (\ref{eta_1}) into the Y-system
(\ref{y-system}) have the properties (\ref{prpty1}), (\ref{prpty2})
up to $n=30$. We have checked for a substantial number of values of
the chemical potential $h$ that they have the third property
(\ref{prpty3}) up to $n=10$.
\item{} Our expression (\ref{eta_1}) agrees with the expansion 
(\ref{fifth_order}) in powers of $\tau$ up to fifth order.
\item{}
Eqn (\ref{eta_1}) agrees perfectly with our numerical solution of the
saddle-point equations discussed in section \ref{numerics}, as is shown
in Fig.~\ref{distributions}. 
\end{enumerate}
Given $\eta_1(x)$ we can use the $Y$-system (\ref{y-system}) to generate
$\eta_{n}(x)$ with $n\geq 2$
\bea
\eta_{n}(x)=\frac{\eta_{n-1}\left(x+\frac{i}{2}\right)\eta_{n-1}\left(x-\frac{i}{2}\right)}{1+\eta_{n-2}(x)}-1\ , \ n\geq 2.
\label{relation1}
\eea
As mentioned before, the distribution functions $\rho_{n}(x)$ can
be obtained using (\ref{rho_n}). The explicit expressions for
$\rho_1(x)$ and $\rho_2(x)$ are as follows:
\bea
 \rho_{1}(x)=\frac{2 \tau^2 (1 + x^2) (1 + 2 \tau + x^2)}{\pi (x^2 + (2 \tau + x^2)^2) (1 + 
   5 x^2 + 4 (\tau + 3 \tau^2 + 4 \tau x^2 + x^4))},
\eea
\bea
\rho_{2}(x)=\frac{16\tau^4(9+4x^2)h_1(x,\tau)}{\pi(1+4x^2+8\tau)h_2(x,\tau)h_3(x,\tau)},
\label{eq:rho_2}
\eea
where
\bea
 h_1(x,\tau)&=&9 + 49 x^2 + 56 x^4 + 16 x^6 + 72 \tau \nonumber\\
 &+& 168 x^2 \tau +  96 x^4 \tau + 116 \tau^2 + 176 x^2 \tau^2 + 96 \tau^3\,,\\
 h_2(x,\tau)&=&9 + 49 x^2 + 56 x^4 + 16 x^6 + 24 \tau \nonumber \\
 &+& 120 x^2 \tau + 
 96 x^4 \tau + 40 \tau^2 + 160 x^2 \tau^2 + 64 \tau^3\,,\\
 h_3(x,\tau)&=&9 x^2 + 49 x^4 + 56 x^6 + 16 x^8 + 96 x^2 \tau + 224 x^4 \tau \nonumber\\
 &+& 
 128 x^6 \tau + 232 x^2 \tau^2 + 352 x^4 \tau^2 + 
 384 x^2 \tau^3 + 144 \tau^4\,.
\eea
The functions $\rho_n(x)$ for $n\geq 3$ are always written as rational functions but their expressions get lengthier as $n$ increases. 

\section{Physical properties of the stationary state}
\label{phys_prop}
\subsection{Local pair correlation function}
\label{section_g2}
The distribution functions $\rho_n(\lambda)$, $\rho^h_n(\lambda)$
completely characterize the stationary state. Their knowledge, in principle,
allows one to calculate all local correlation functions in the thermodynamic
limit. In practice, while formulas exist for the
expectation values of simple local operators in the Lieb-Liniger model
in the finite volume \cite{slavnov, ccs-07, pozsgay-11, pc-15}, it is
generally difficult to take the thermodynamic limit of these expressions. 
In contrast to the repulsive case\cite{pozsgay-11, gs-03, kgds-03,
  kmt-09, kci-11, nrtg-16}, much less is known in the attractive
regime, where technical complications arise that are associated with
the existence of string solutions to the Bethe ansatz equations. Here
we focus on the computation of the local pair correlation function
\be
g_2=\frac{\langle :\hat{\rho}^2(0):\rangle}{D^2}=\frac{\langle \Psi^{\dagger}(0)\Psi^{\dagger}(0)\Psi(0)\Psi(0)\rangle}{{D^2}}.
\label{definition_g2}
\ee

We start by applying the Hellmann-Feynman \cite{gs-03, kgds-03,
  kci-11,mp-14} theorem to the expectation value in a general
energy eigenstate $|\{\lambda_j\}\rangle$ with energy $E[\{\lambda_j\}]$
of the finite system
\be
\langle \{\lambda_j\}| \Psi^{\dagger}\Psi^{\dagger}\Psi\Psi|\{\lambda_j\}\rangle=-\frac{1}{L}\frac{\partial E[\{\lambda_j\}]}{\partial \overline{c}}\ .
\label{hell_fey}
\ee
In order to evaluate the expression on the r.h.s., we need to take the
derivative of the Bethe-Takahashi equations (\ref{BGT})
with respect to $\overline{c}$
\begin{equation}
f^{(n)}(\lambda_{\alpha})=\frac{1}{n}\sum_{m}\frac{2\pi}{L}\sum_{\beta}\left(f^{(n)}(\lambda_{\alpha})-f^{(m)}(\lambda_{\beta})-\frac{\lambda^{n}_{\alpha}}{\overline{c}}+\frac{\lambda^{m}_{\beta}}{\overline{c}}\right)a_{nm}(\lambda^{n}_{\alpha}-\lambda^{m}_{\beta})\ .
\end{equation}
Here $a_{nm}$ is given in Eq. (\ref{aa_function}) and
\begin{equation}
f^{(n)}(\lambda_{\alpha})=\frac{\partial\lambda^{n}_{\alpha}}{\partial \overline{c}}\ .
\end{equation}
Taking the thermodynamic limit gives
\begin{eqnarray}
f^{(n)}(\lambda)=\frac{2\pi}{n}\left(f^{(n)}(\lambda)-\frac{\lambda}{\overline{c}}\right)\sum_{m=1}^{\infty}\int_{-\infty}^{\infty} \mathrm{d}\mu\ \rho_{m}(\mu)a_{nm}(\lambda-\mu)&\ \nonumber \\
+ \frac{2\pi}{n}\sum_{m=1}^{\infty}\int_{-\infty}^{\infty}\mathrm{d}\mu\ \rho_{m}(\mu)\left(\frac{\mu}{\overline{c}}-f^{(m)}(\mu)\right)a_{nm}(\lambda-\mu).&\ 
\end{eqnarray}
Using the thermodynamic version of the Bethe-Takahashi equations
(\ref{coupled}) and defining 
\begin{equation}
b_{n}(\lambda)=2\pi\left(\frac{\lambda}{\overline{c}}-f^{(n)}(\lambda)\right)\rho_{n}^{t}(\lambda),
\label{a:b_function}
\end{equation}
we arrive at
\begin{equation}
b_{n}(\lambda)=n\frac{\lambda}{\overline{c}}-\sum_{m=1}^{\infty}\int_{-\infty}^{\infty}\mathrm{d}\mu\ \frac{1}{1+\eta_{m}(\mu)}b_{m}(\mu)a_{nm}(\lambda-\mu)\ .
\label{a:aux_1}
\end{equation}
The set of equations (\ref{a:aux_1}) completely fixes the functions
$b_{n}(\lambda)$, once the functions $\eta_n(\lambda)$ are
known. The right hand side of (\ref{hell_fey}) in the finite volume
can be cast in the form
\be
\frac{\partial E}{\partial \overline{c}}=\sum_{n}\left[\sum_{\alpha}2n\lambda_{\alpha}^{n}f^{(n)}(\lambda_{\alpha})-\frac{\overline{c}}{6}n(n^2-1)\right]\ .
\ee
Taking the thermodynamic limit, and using (\ref{a:b_function}) we finally arrive at
\be
\frac{1}{L}\frac{\partial E}{\partial \overline{c}}=\sum_{n=1}^{\infty}\int_{-\infty}^{\infty}\frac{\mathrm{d}\mu}{2\pi}\ \left[2\pi \rho_{n}(\mu)\left(\frac{2n\mu^2}{\overline{c}}-\frac{\overline{c}}{6}n(n^2-1)\right)-2n\mu b_n(\mu)\frac{1}{1+\eta_{m}(\mu)}\right].
\label{a:aux_2}
\ee
Combining (\ref{a:aux_1}) and (\ref{a:aux_2}) we can express the local
pair correlation function as
\be
g_2(\gamma)=\gamma^2\sum_{m=1}^{\infty}\int_{-\infty}^{\infty}\frac{\mathrm{d}x}{2\pi}\ \left[2mxb_m(x)\frac{1}{1+\widetilde{\eta}_m(x)}  - 2\pi \widetilde{\rho}_m(x)\left(2mx^2-\frac{m(m^2-1)}{6}\right)\right] ,
\label{one}
\ee
where the functions $b_{n}(x)$ are determined by
\be
 b_n(x)=nx-\sum_{m=1}^{\infty}\int_{-\infty}^{\infty}\mathrm{d}y\ \frac{1}{1+\widetilde{\eta}_{m}(y)}b_{m}(y)\widetilde{a}_{nm}(x-y).
\label{two}
\ee
In (\ref{one}), (\ref{two}) we defined
\be
\widetilde{\eta}_{n}(x)=\eta_{n}(x\overline{c})\ ,\quad
\widetilde{\rho}_{n}(x)=\rho_{n}(x\overline{c})\ ,\quad
\widetilde{a}_{nm}(x)=\overline{c}a_{nm}(x\overline{c}).
\ee
Using the
knowledge of the functions $\eta_n(\lambda)$ for the
stationary state, we can solve Eqns~(\ref{two}) numerically and
substitute the results into (\ref{one}) to obtain $g_2(\gamma)$. 

While (\ref{one}), (\ref{two}) cannot be solved in closed form, they
can be used to obtain an explicit asymptotic expansion around
$\gamma=\infty$. To that end we use (\ref{therm_momentum_energy}),
(\ref{eq:epsilon}) and (\ref{energy}) to rewrite  $g_2(\gamma)$ as 
\be
g_2(\gamma)=2+\gamma^2\sum_{m=1}^{\infty}\int_{-\infty}^{\infty}\frac{\mathrm{d}x}{2\pi}2mxb_m(x)\frac{1}{1+\widetilde{\eta}_m(x)}.
\label{new_g2}
\ee
We then use that large values of $\gamma$ correspond to small values
of $\tau$, cf. (\ref{fact}), and carry out a small-$\tau$ expansion of
the functions
\be
\frac{1}{1+\widetilde{\eta}_n(x)}=\frac{\widetilde{\varphi}_n(x)}{(1+\widetilde{\varphi}_n(x))},
\label{expansion_2}
\ee
where $\widetilde{\varphi}_n(x)=1/\widetilde{\eta}_n(x)$ as in
section~\ref{perturbative_sec}. Substituting this expansion into the
r.h.s. of (\ref{two}) and proceeding iteratively, we obtain an
expansion for the functions $b_n(x)$ in powers of $\tau$. The steps
are completely analogous to those discussed in
section~\ref{perturbative_sec} for the functions $\varphi_n(\lambda)$  
and will not be repeated here. Finally, we use the series expansions of
$b_n(x)$ and $(1+\widetilde{\eta}_n(x))^{-1}$ in (\ref{new_g2}) to
obtain an asymptotic expansion for $g_2(\gamma)$. The result is
\be
g_{2}(\gamma)=4-\frac{40}{3\gamma}+\frac{344}{3\gamma^2}-\frac{2656}{3\gamma^3}+\frac{1447904}{225\gamma^4}+\mathcal{O}(\gamma^{-5}).
\label{analytical_g2}
\ee
\begin{figure}[ht]
\centering
\includegraphics[scale=0.95]{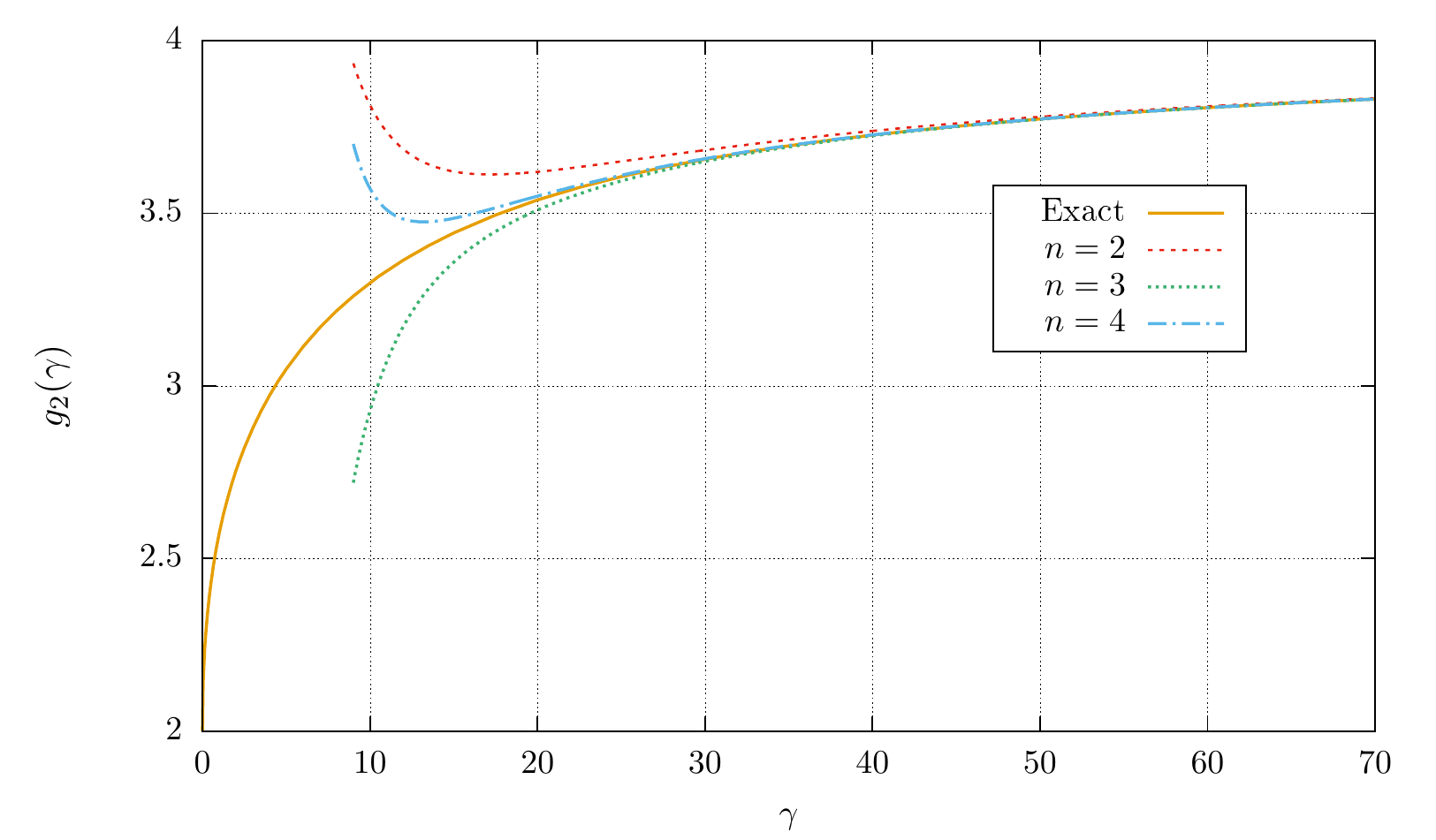}
\caption{Local pair correlation function $g_2(\gamma)$ in the
stationary state at late times after the quench. The numerical solution
of Eqns (\ref{one}), (\ref{two}) is shown as a solid orange line. The
asymptotic expansion (\ref{analytical_g2}) around $\gamma=\infty$
up to order $\mathcal{O}(\gamma^{-n})$ with $n=2,3,4$ is seen to be in
good agreement for large values of $\gamma$.} 
\label{local_pair}
\end{figure}
In Fig. \ref{local_pair} we compare results of a full numerical
solution of Eqns (\ref{one}), (\ref{two}) to the asymptotic expansion
(\ref{analytical_g2}). As expected, the latter breaks down for
sufficiently small values of $\gamma$. In contrast to the
large-$\gamma$ regime, the limit $\gamma\to 0$ is more difficult to
analyze because $g_2(\gamma)$ is non-analytic in $\gamma=0$. 
The limit $\gamma\to 0$ can be calculated as shown in
Appendix~\ref{small_gamma}, and is given by
\be
\lim_{\gamma\to 0}g_2(\gamma)=2.
\label{limit_0}
\ee
As was already noted in Ref.~\cite{pce-16}, (\ref{limit_0}) implies
that the function $g_2(\gamma)$ is discontinuous in
$\gamma=0$. Indeed, $g_2(0)$ can be calculated directly by using
Wick's theorem in the initial BEC state
\be
\frac{\langle {\rm BEC}|:\hat{\rho}(0)^2:|{\rm BEC}\rangle}{D^2} = 1.
\ee
This discontinuity, which is absent for quenches to the repulsive
regime \cite{dwbc-14}, is ascribed to the presence of multi-particle
bound states for all values of $\gamma\neq 0$. The former are also at
the origin of the non-vanishing limit of $g_2(\gamma)$ for
$\gamma\to\infty$ as it will be discussed in the next section. 

Finally, an interesting question is the calculation of the three-body one-point correlation function $g_3(\gamma)$ on the post-quench steady state. The latter is relevant for experimental realizations of bosons confined in one dimension, as it is proportional to the three-body recombination rate \cite{lohp-04}. For $g_3$ it is reasonable to expect that three-particle bound states may give non-vanishing contributions in the large coupling limit. 
While $g_3$ is known for general states in the repulsive Lieb-Liniger model, its computation in the attractive case is significantly harder and requires further development of existing methods. We hope that our work will motivate theoretical efforts in this direction.

\subsection{Physical implications of the multi-particle bound states}
\label{physical_discussions}
\begin{figure}[ht]
\centering
\includegraphics[scale=0.82]{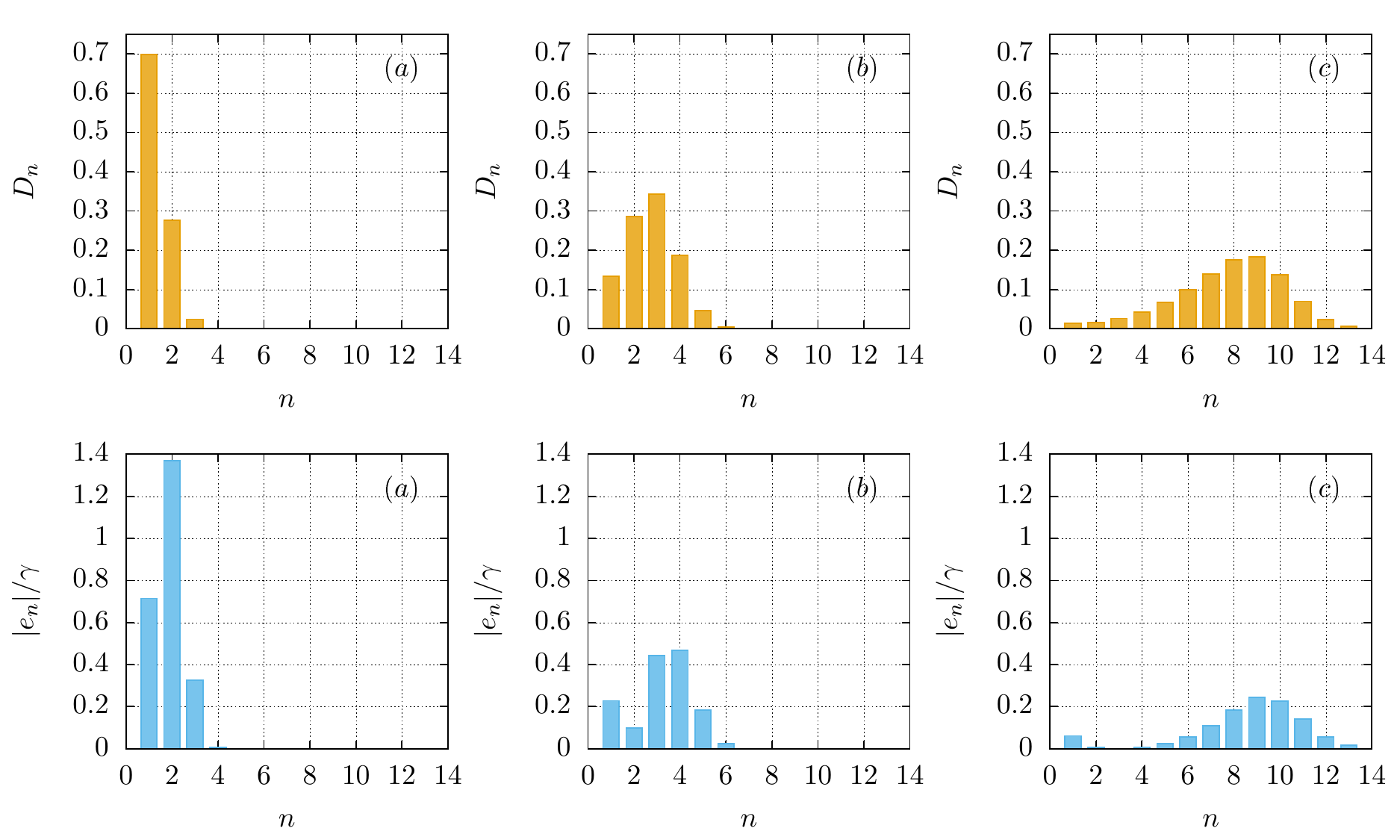}
\caption{ Density $D_n$ and absolute value of the normalized energies
per volume $e_n/\gamma$ of the bosons forming $n$-particle bound
states as defined in (\ref{density_per_string}). The plots correspond to
$(a)$ $\gamma=20$, $(b)$ $\gamma=2$, $(c)$ $\gamma=0.2$. The total
density is fixed $D=1$. The energy densities $e_n$ are always negative
for $n\geq 2$ (i.e. $|e_n|=-e_n$ for $n\geq 2$) while $e_1>0$.} 
\label{strings}
\end{figure}

A particularly interesting feature of our stationary state is the
presence of finite densities of $n$-particle bound states with 
$n\geq 2$. In Fig.~\ref{strings}, their densities and energies per
volume are shown for a number of different values of $\gamma$.  
We see that the maximum of $D_n$ occurs at a value of $n$ that
increases as $\gamma$ decreases. This result has a simple physical
interpretation. In the attractive regime, the bosons have a tendency
to form multi-particle bound states. One might naively expect that
increasing the strength $\gamma$ of the attraction between bosons
would lead to the formation of bound states with an ever increasing
number of particles, thus leading to phase separation. However, in the
quench setup the total energy of the system is fixed by the initial
state, cf. (\ref{energy}), while the energy of $n$-particle
bound states scales as $n^3$, cf. Eqns~(\ref{eq:epsilon}),
(\ref{density_per_string}). As a result, $n$-particle bound states
cannot be formed for large values of $\gamma$, and indeed they are
found to have very small densities for $n\geq 3$. On the contrary,
decreasing  the interaction strength $\gamma$, the absolute value of
their energy lowers and these bound states become accessible. The
dependence of the peak in Fig.~\ref{strings} on $\gamma$ is monotonic
but non-trivial and it is the result of the competition between the
tendency of attractive bosons to cluster, and the fact that
$n$-particle bound states with $n$ very large cannot be formed as a
result of energy conservation.

The presence of multi-particle bound states affects measurable
properties of the system, and is the reason for the particular
behaviour of the local pair correlation function computed in the
previous section. Remarkably, this is true also in the limit
$\gamma\to \infty$. This is in marked contrast to the super
Tonks-Girardeau gas, where bound states are absent. To exhibit the
important role of bound states in the limit of large $\gamma$, we will
demonstrate that the limiting value of $g_2(\gamma)$ for $\gamma\to
\infty$ is entirely determined by bound pairs. It follows from
(\ref{one}) that $g_2(\gamma)$ can be written in the form
\be
g_2(\gamma)=\sum_{m=1}^{\infty}g_{2}^{(m)}(\gamma),
\ee
where $g_2^{(m)}(\gamma)$ denotes the contribution of
$m$-particle bound states to the local pair correlation
\be
g_2^{(m)}(\gamma)= \gamma^2\int_{-\infty}^{\infty}\frac{\mathrm{d}x}{2\pi}\ \left[2mxb_m(x)\frac{1}{1+\widetilde{\eta}_m(x)}  - 2\pi \widetilde{\rho}_m(x)\left(2mx^2-\frac{m(m^2-1)}{6}\right)\right].
\label{eq:temp_1}
\ee
Let us first show that unbound particles give a vanishing contribution
\be
\lim_{\gamma \to \infty}g_{2}^{(1)}(\gamma)=0.
\label{limit1}
\ee
In order to prove this, we use that at leading order in $1/\gamma$ we
have $b_1(x)=x$. Using the explicit expressions for
$\widetilde{\eta}_1(x)$, $\widetilde{\rho}_1(x)$  we can then perform
the integrations in the r.h.s. of Eq.~(\ref{eq:temp_1}) exactly and
take the limit $\gamma\to \infty$ afterwards. We obtain  
\bea
\lim_{\gamma\to\infty}\gamma^2\int_{-\infty}^{\infty}\frac{\mathrm{d}x}{2\pi}\ 2xb_1(x)\frac{1}{1+\widetilde{\eta}_1(x)}=2,\\
\lim_{\gamma\to\infty}\gamma^2\int_{-\infty}^{\infty}\frac{\mathrm{d}x}{2\pi}\ \left(- 2\pi \widetilde{\rho}_1(x)2x^2\right)=-2,
\eea
which establishes (\ref{limit1}). Next, we address the bound pair
contribution. At leading order in $1/\gamma$ we have $b_2(x)=2x$, and
using the explicit expression for $\widetilde{\eta}_{2}(x)$ we obtain 
\be
\lim_{\gamma\to\infty}\gamma^2\int_{-\infty}^{\infty}\frac{\mathrm{d}x}{2\pi}\ 4xb_2(x)\frac{1}{1+\widetilde{\eta}_2(x)}=0.
\ee
This leaves us with the contribution
\be
\lim_{\gamma\to \infty} \gamma^2\int_{-\infty}^{\infty}\frac{\mathrm{d}x}{2\pi}\ \left[ -2\pi \widetilde{\rho_2}(x)\left(4x^2-1\right)\right].
\label{limit_2}
\ee
Although the function $\widetilde{\rho}_2(x)$ is known,
cf. Eq.~(\ref{eq:rho_2}), its expression is unwieldy and it is
difficult to compute the integral analytically. On the other hand, one
cannot expand $\widetilde{\rho}_2(x)$ in $1/\gamma$ inside the
integral, because the integral of individual terms in this expansion
are not convergent (signalling that in this case one cannot exchange
the order of the limit $\gamma\to\infty$ and of the
integration). Nevertheless, the numerical computation of the integral
in (\ref{limit_2}) for large values of $\gamma$ presents no
difficulties and one can then compute the limit numerically. We found
that the limit in Eq.~(\ref{limit_2}) is equal to $4$ within machine
precision so that 
\be
\lim_{\gamma\to\infty} g_2(\gamma)=4= \lim_{\gamma\to\infty} g_2^{(2)}(\gamma).
\ee
Finally, we verified that contributions coming from bound states
with higher numbers of particles are vanishing, i.e.
$g_2^{(m)}(\gamma)\to 0$ for $\gamma\to \infty$, $m\geq 3$. This establishes
that the behaviour of $g_2(\gamma)$ for large values of $\gamma$ is
dominated by bound pair of bosons.

\section{Conclusions}
\label{conclusions}
We have considered quantum quenches from an ideal Bose condensate to
the one-dimensional Lieb-Liniger model with arbitrary attractive
interactions. We have determined the stationary state,
and determined its physical properties. In particular, we revealed
that the stationary state is composed of an interesting mixture of
multi-particle bound states, and computed the local pair
correlation function in this state. Our discussion presents a detailed
derivation of results first announced in Ref.~\cite{pce-16}. 

As we have stressed repeatedly, the most intriguing feature of the
stationary state for the quench studied in this work is the presence
of multi-particle bound states. As was argued in Ref.~\cite{pce-16},
their properties could in principle be probed in ultra-cold atoms
experiments. Multi-particle bound states are also formed in the quench
from the N\'eel state to the gapped XXZ model, as it was recently
reported in Refs.~\cite{wdbf-14,bwfd-14,budapest,buda2}. However, in
contrast to our case, the bound state densities are always small
compared to the density of unbound magnons for all the values of the
final anisotropy parameter $\Delta\geq 1$ \cite{bwfd-14}.

Our work also provides an interesting physical example of a quantum
quench, where different initial conditions lead to stationary states
with qualitatively different features. Indeed, a quench in the
one-dimensional Bose gas from the infinitely repulsive to the
infinitely attractive regime leads to the super Tonks-Girardeau 
gas, where bound states are absent. On the other hand, as shown in
section \ref{phys_prop}, if the initial state is an ideal Bose
condensate,  bound states have important consequences on the
correlation functions of the system even in the limit of large
negative interactions.  

An interesting open question is to find a description of our
stationary state in terms of a GGE. As the stationary state involves
bound states, it is likely that the GGE will involve not yet known
quasi-local conserved charges \cite{idwc-15,iqdb-15,impz-16} as well as the
known ultra-local ones\cite{davies-90}. In the Lieb-Liniger model
technical difficulties arise when addressing such issues, as
expectation values of local conserved charges generally exhibit divergences
\cite{davies-90,kscc-13, kcc-14}. In addition, very little is known
about quasi-local conserved charges for interacting models defined in
the continuum \cite{impz-16, emp-15}.

Finally, it would be interesting to investigate the approach to the steady state in the quench considered in this work. While this is in general a very difficult problem, in the repulsive regime the post-quench time evolution from the non-interacting BEC state was considered in \cite{dpc-15}. There an efficient numerical evaluation of the representation \eqref{time_ev} was performed, based on the knowledge of exact one-point form factors \cite{pc-15}. The attractive regime, however, is significantly more involved due to the presence of bound states and the study of the whole post-quench time evolution remains a theoretical challenge for future investigations.

\section*{Acknowledgements}
We thank  Michael Brockmann for a careful reading of the manuscript. PC acknowledges the financial support by the ERC under Starting Grant
279391 EDEQS. The work of FHLE was supported by the EPSRC under grant
EP/N01930X. All authors acknowledge the hospitality of the Isaac
Newton Institute for Mathematical Sciences under grant EP/K032208/1.


\begin{appendix}

\section{Overlaps in the presence of zero-momentum $n$-strings}
\label{app_overlap}
In this appendix we argue that Eq. (\ref{leading}) gives the leading
term in the thermodynamic limit of the logarithm of the overlap
between the BEC state and a parity-invariant Bethe state, even in
cases where the latter contains zero-momentum strings.   

To see this, consider a parity invariant Bethe state 
with a single zero-momentum $m$-string, and $K$ parity-related pairs
of $n_j$-strings. The total number of particles in such a state is then
$N=2\sum_{j}n_j+m$. In Ref.~\cite{cl-14} an explicit expression for
the overlap (\ref{overlap}) of such states with a BEC state in the
zero-density limit ($L\to \infty$ and $N$ fixed) was obtained. Up to an 
irrelevant (for our purposes) overall minus sign, it reads 
\bea
\langle \{\lambda_{j}\}_{j=1}^{N/2}\cup \{-\lambda_{j}\}_{j=1}^{N/2} |{\rm BEC}\rangle &=& \frac{2^{m-1}L\overline{c}}{(m-1)!}\sqrt{\frac{N!}{(L\overline{c})^{N}}}\nonumber\\
&\times&\prod_{p=1}^{K}\frac{1}{\sqrt{\frac{\lambda_p^2}{\overline{c}^2}\left(\frac{\lambda_p^2}{\overline{c}^2}+\frac{n_p^2}{4}\right)}\prod_{q=1}^{n_p-1}\left(\frac{\lambda_p^2}{\overline{c}^2}+\frac{q^2}{4}\right)},
\label{zero_momentum_overlap}
\eea 
where $\lambda_p$ is the centre of the $p$'th string. We see that as a
result of having a zero-momentum string, an additional pre-factor $L$
appears. In general, the presence of $M$ zero-momentum strings will
lead to an additional pre-factor $L^M$ \cite{cl-14}. While
(\ref{zero_momentum_overlap}) is derived in the zero density limit,
we expect an additional pre-factor to be present also if one considers
the thermodynamic limit $N,L\to \infty$, at finite density $D=N/L$.
Importantly such pre-factors will result in \emph{sub-leading}
corrections of order $(\ln L)/ L$ to the logarithm of the
overlaps. This suggests that (\ref{leading}) holds even for states with
zero-momentum $n$-strings.  

\section{Tri-diagonal form of the coupled integral equations}
\label{app_tridiag}
\subsection{Tri-diagonal Bethe-Takahashi equations}
Our starting point are the thermodynamic Bethe equations
(\ref{coupled}). For later convenience we introduce the following
notations for the Fourier transform of a function
\begin{equation}
\hat{f}(k)=\mathcal{F}[f](k)=\int_{-\infty}^{\infty}f(\lambda)e^{ik\lambda}\mathrm{d}\lambda\ ,
\end{equation}
\begin{equation}
f(\lambda)=\mathcal{F}^{-1}[\hat{f}](\lambda)=\frac{1}{2\pi}\int_{-\infty}^{\infty}\hat{f}(k)e^{-ik\lambda}\mathrm{d}k\ .
\end{equation}
We recall that $f\ast g$ denotes the convolution of two functions,
cf. (\ref{convolution}). The Fourier transform of $a_{n}(\lambda)$
defined in (\ref{a_function}) is easily computed 
\begin{equation}
\hat{a}_{n}(k)=e^{-\frac{n\overline{c}|k|}{2}}\ .
\end{equation}
Following Ref. \cite{gaudin}, we introduce the symbols
\begin{equation}
[nmp]=\left\{\begin{array}{cc}1\ , & \mathrm{if}\ p=|m-n|\ \mathrm{or}\ m+n\\2\ , & \mathrm{if}\  p= |m-n|+2,\ |m-n|+4, \ldots, m+n-2\ , \\0 & \mathrm{otherwise}\ .\end{array}\right.
\end{equation}
We can then perform the Fourier transform of both sides of (\ref{coupled}) and obtain
\begin{equation}
n\delta(k)-\sum_{m=1}\sum_{p>0}[nmp]\hat{\rho}_{m}(k)e^{-\frac{\overline{c}}{2}|k|p}=\hat{\rho}_n^t(k)\ ,
\label{a:intermediate}
\end{equation}
where $\rho_n^t(\lambda)$ are given in (\ref{rho_tot}). We now define
\begin{eqnarray}
\hat{\rho}_{-m}(k)=-\hat{\rho}_{m}(k)\ ,\qquad m\geq 1 ,&\\
\hat{\rho}_{0}(k)=0\ . &
\end{eqnarray}
After straightforward calculations, we can rewrite (\ref{a:intermediate}) in the form
\begin{equation}
\hat{\rho}_{n}^h(k)=n\delta(k)-\coth\left(\frac{|k|\overline{c}}{2}\right)\sum_{m=-\infty}^{+\infty}e^{-|k||n-m|\frac{\overline{c}}{2}}\hat{\rho}_{m}(k)\ .
\label{temp_1}
\end{equation}
In order to decouple these equations we note that
\begin{eqnarray}
\hat{\rho}_{n+1}^{h}(k)&+&\hat{\rho}_{n-1}^{h}(k)=2n\delta(k)\nonumber \\
&-&\coth\left(\frac{|k|\overline{c}}{2}\right)\left[-2\hat{\rho}_{n}(k)\sinh\left(\frac{|k|\overline{c}}{2}\right)+2\cosh\left(\frac{|k|\overline{c}}{2}\right)\sum_{m=-\infty}^{\infty}e^{-|k||n-m|\frac{\overline{c}}{2}}\hat{\rho}_{m}(k)\right]\ .\label{temp_2}
\end{eqnarray}
Combining Eqns (\ref{temp_1}), (\ref{temp_2}) one obtains
\begin{eqnarray}
\hat{\rho}_{n}^t(k)&=&\frac{1}{2\cosh\left(|k|\overline{c}/2\right)}\left(\hat{\rho}_{n+1}^h(k)+\hat{\rho}_{n-1}^h(k)\right)-\underbrace{n\delta(k)\left[\frac{1-\cosh\left(\frac{|k|\overline{c}}{2}\right)}{\cosh\left(\frac{|k|\overline{c}}{2}\right)}\right]}_{=0}=\nonumber\\
&=&\frac{1}{2\cosh\left(|k|\overline{c}/2\right)}\left(\hat{\rho}_{n+1}^h(k)+\hat{\rho}_{n-1}^h(k)\right)\ .
\end{eqnarray}
We can now perform the inverse Fourier transform. Using
\begin{equation}
\frac{1}{2\pi}\int_{-\infty}^{\infty}\ d k\frac{1}{\cosh\left(k\frac{\overline{c}}{2}\right)}e^{-i\lambda k}=\frac{1}{\overline{c}}\frac{1}{\cosh\left(\frac{\lambda\pi}{\overline{c}}\right)}\ ,
\label{a:integral}
\end{equation}
we finally obtain
\begin{eqnarray}
				\rho_{n}(1+\eta_{n})=s\ast\left(\eta_{n-1}\rho_{n-1}+\eta_{n+1}\rho_{n+1}\right)\qquad n\geq 1\ , \label{a:final_bethe}
\label{a:final_bethe_2}	
\end{eqnarray}
where we can choose $\eta_{0}(\lambda)\rho_0(\lambda)=\delta(\lambda)$, $\eta_n(\lambda)$ is given in Eq.~(\ref{eq:eta}), and where 
\begin{equation}
s(\lambda)=\frac{1}{2\overline{c}\cosh\left(\frac{\pi\lambda}{\overline{c}}\right)}\ .
\label{a:kernel}
\end{equation}
\subsection{Tri-diagonal oTBA equations}
In this appendix we derive the tri-diagonal equations
(\ref{finale_gtba}) starting from Eqns (\ref{coupled2}). Our
discussion follows Ref.~\cite{wdbf-14}. Some useful identities are\cite{takahashi} 
\begin{equation}
(a_0+a_2)\ast a_{nm}=a_1\ast(a_{n-1,m}+a_{n+1,m})+(\delta_{n-1,m}+\delta_{n+1,m})a_1\ ,\qquad n\geq 2,\ m\geq 1,
\end{equation}
\begin{equation}
(a_0+a_2)\ast a_{1m}=a_1\ast a_{2,m}+a_1\delta_{2,m}\ , \qquad m\geq 1\ ,
\end{equation}
where we define $a_0(\lambda)=\delta(\lambda)$, and where the
functions $a_{nm}(\lambda)$, $a_n(\lambda)$ are given in
Eqns~(\ref{aa_function}), (\ref{a_function}). Convolution of
(\ref{coupled2}) with $(a_0+a_2)$ gives
\begin{eqnarray}
(a_0+a_2)\ast\ln\eta_n&=&(a_0+a_2)\ast g_n-a_1\ast (g_{n-1}+g_{n+1})\nonumber \\
&+&a_{1}\ast\left[\ln(1+\eta_{n-1})+\ln(1+\eta_{n+1})\right]\ ,\qquad n\geq 1\ ,
\label{a:simplified}
\end{eqnarray}
where we defined $g_{n}(\lambda)=-\ln W_n(\lambda)$, $g_0(\lambda)=0$ and
$\eta_0(\lambda)=0$. The functions $g_n(\lambda)$ can be written as 
\begin{equation}
g_{n}(\lambda)=\ln s_{0}^{(2)}(\lambda)+\ln s_{n}^{(2)}(\lambda)+2\sum_{\ell=1}^{n-1}\ln s_{\ell}^{(2)}(\lambda)\ ,
\label{a:gn}
\end{equation}
where
\begin{equation}
s_{\ell}^{(2)}(\lambda)=s_{\ell}(\lambda)s_{-\ell}(\lambda)=\frac{\lambda^2}{\overline{c}^2}+\frac{\ell^2}{4}\ .
\end{equation}
It is straightforward to show that
\begin{equation}
(a_{m}\ast f_{r})(\lambda)=f_{m+r}(\lambda)\,,
\label{a:identity}
\end{equation}
where we defined
\be
f_{r}(\lambda)=\ln\left[\left(\frac{\lambda}{\overline{c}}\right)^2+\left(\frac{r}{2}\right)^2\right]\,.
\label{ffunction}
\ee
Using (\ref{a:identity}) and (\ref{a:gn}), we can rewrite the driving
term in (\ref{a:simplified}) as 
\begin{eqnarray}
\tilde{d}_n\equiv (a_0+a_{2})\ast g_n-a_1\ast (g_{n-1}+g_{n+1})=f_{0}-f_{2}=\ln\left(\frac{\lambda^2}{\overline{c}^2}\right)-\ln\left(\frac{\lambda^2}{\overline{c}^2}+1\right)\ ,&
\label{a:important1}
\end{eqnarray}
which allows us to rewrite the oTBA equations in the form
\begin{equation}
(a_0+a_2)\ast \ln\eta_n=\tilde{d}_n+a_{1}\ast\left[\ln(1+\eta_{n-1})+\ln(1+\eta_{n+1})\right]\ .
\label{a:important2}
\end{equation}
We note that $\tilde{d}_n$ is in fact independent of $n$. Carrying out the
Fourier transform and using that $f_0-f_2=(a_0-a_2)\ast f_0$ we obtain
\begin{eqnarray}
\mathcal{F}\left[\ln \eta_n\right]&=&\frac{1}{1+e^{-\overline{c}|k|}}(1-e^{-\overline{c}|k|})\mathcal{F}\left[f_0\right]\nonumber \\
&+&\frac{1}{1+e^{-\overline{c}|k|}}e^{-\frac{\overline{c}|k|}{2}}\mathcal{F}\left[\left(\ln(1+\eta_{n-1})+\ln(1+\eta_{n+1})\right)\right]\ .
\label{a:aa}
\end{eqnarray}
The first term on the right hand side simplifies
\begin{equation}
\frac{1}{1+e^{-\overline{c}|k|}}(1-e^{-\overline{c}|k|})\mathcal{F}\left[f_0\right]=-2\pi\frac{\tanh(\overline{c}k/2)}{k}\ .
\end{equation}
Finally, taking the inverse Fourier transform of (\ref{a:aa}), using
(\ref{a:integral}) as well as 
\begin{equation}
\int_{-\infty}^{\infty} d k e^{-i k \lambda} \frac{\tanh(\overline{c}k/2)}{k}=-\ln\left[\tanh^2\left(\frac{\pi\lambda}{2\overline{c}}\right)\right]\ ,
\end{equation}
we arrive at the desired tri-diagonal form of the oTBA equations
\begin{eqnarray}
      \ln(\eta_n)=d+s\ast\left[\ln(1+\eta_{n-1})+\ln(1+\eta_{n+1})\right]\ ,\qquad n\geq 1\ , &\\
      \eta_{0}(\lambda)=0\ .&
\label{a:finale gtba}	
\end{eqnarray}
Here $s(\lambda)$ is given by Eq. (\ref{a:kernel}) and
\be
d(\lambda)=\ln\left[\tanh^2\left(\frac{\pi\lambda}{2\overline{c}}\right)\right]. \ee

\section{Asymptotic behaviour}
\label{app_asymptotic}
In this appendix we derive the asymptotic condition (\ref{difference})
for the tri-diagonal equations (\ref{finale_gtba}). Our derivation closely
follows the finite temperature case \cite{takahashi}. We start from
Eq.~(\ref{coupled2}) for $n=1$
\be
\ln \eta_{1}(\lambda)=-2h+(f_0+f_1)+a_2\ast\ln(1+\eta^{-1}_{1})+\sum_{m=2}^{+\infty}(a_{m-1}+a_{m+1})\ast\ln\left(1+\eta^{-1}_{m}\right)\ ,
\label{case1}
\ee
where $f_{r}=f_{r}(\lambda)$ is defined in (\ref{ffunction}). We use now the following identities, which are easily derived from (\ref{a:identity}), (\ref{a:important1}), (\ref{a:important2})
\begin{eqnarray}
a_2\ast \ln (1+\eta_{1}^{-1})&=&a_2\ast \ln(1+\eta_1)-a_2\ast \ln \eta_1=\nonumber \\
&=&a_2\ast \ln(1+\eta_1)-f_0+f_2-a_1\ast \ln (1+\eta_2)+\ln \eta_1\ .
\label{passaggio}
\end{eqnarray}
Using (\ref{passaggio}) we can recast (\ref{case1}) in the form
\begin{eqnarray}
-2h+a_1\ast(f_0+f_1)=a_1\ast \ln \eta_2&-&a_2\ast\ln(1+\eta_1)-a_3\ast\ln(1+\eta_2^{-1})\nonumber \\
 &-&\sum_{m=3}^{+\infty}(a_{m-1}+a_{m+1})\ast\ln(1+\eta_m^{-1})\ .
\label{eq:aaa}
\end{eqnarray}
To proceed, we write
\bea
\sum_{m=3}^{+\infty}(a_{m-1}+a_{m+1})\ast\ln(1+\eta_m^{-1})
=(a_{2}+a_{4})\ast\ln(1+\eta_{3}^{-1})+\sum_{m=4}^{+\infty}(a_{m-1}+a_{m+1})\ast\ln(1+\eta_m^{-1})\ .
\eea
After rewriting the first term on the right hand side, we substitute
back into (\ref{eq:aaa}) to obtain
\bea
-2h+a_2\ast(f_0+f_1)=a_2\ast \ln \eta_3&-&a_3\ast\ln(1+\eta_2)-a_4\ast\ln(1+\eta_3^{-1})\nonumber \\
&-&\sum_{m=4}^{+\infty}(a_{m-1}+a_{m+1})\ast\ln(1+\eta_m^{-1}).
\label{eq:aab}
\eea
Iterating the above procedure $n$ times we arrive at
\bea
-2h+a_n\ast(f_0+f_1)=a_n\ast \ln \eta_{n+1}&-&a_{n+1}\ast\ln(1+\eta_{n})-a_{n+2}\ast\ln(1+\eta_{n+1}^{-1})\nonumber \\
&-&\sum_{m=n+2}^{+\infty}(a_{m-1}+a_{m+1})\ast\ln(1+\eta_m^{-1})\ .
\label{eq:aac}
\eea
Fourier transforming and using the definition for $f_r$ given in
(\ref{ffunction}) we obtain 
\bea
 \ln\eta_{n+1}&=&-2h+\ln\left[\frac{\lambda}{\overline{c}}\left(\frac{\lambda^2}{\overline{c}^2}+\frac{1}{4}\right)\right]+a_1\ast\ln\eta_n\nonumber \\
 &+&a_{1}\ast\ln(1+\eta_{n}^{-1})+a_{2}\ast\ln(1+\eta_{n+1}^{-1})+\sum_{m=2}^{+\infty}(a_{m-1}+a_{m+1})\ast\ln(1+\eta_{m+n}^{-1}).
\label{almost_final}
\eea
Assuming that $\eta_{n}^{-1}(\lambda)$ is vanishing sufficiently fast
as $n\to \infty$ for a generic (and fixed) value of $\lambda$, we can
drop the infinite sum and the two previous terms, and arrive at
Eq.~(\ref{difference}). 
\section{Perturbative analysis}
\label{app_perturbative}
In this appendix we sketch the calculations leading to the expansion (\ref{fifth_order}). Throughout this appendix we work with the dimensionless variable $x=\lambda/\overline{c}$. At the lowest order, it follows from Eq.~(\ref{coefficient}) that 
\be
\varphi_1(x)=\frac{\tau^2}{x^2\left(x^2+\frac{1}{4}\right)}+\mathcal{O}(\tau^3).
\label{second_order}
\ee
Since $\varphi_n(x)\propto \tau^{2n}$, we can neglect $\varphi_n(x)$
with $n\geq 2$ to compute the third order expansion of
$\varphi_1(x)$. Hence, the infinite sum in (\ref{perturbative}) for
$n=1$ can be truncated, at third order in $\tau$, to the first term
($m=1$), where we can use the expansion (\ref{second_order}) for
$\varphi_{1}(\lambda)$. Following Ref.~\cite{dwbc-14} one can then use
identity (\ref{a:identity}) to perform the convolution integral and
finally obtain 
\be
\varphi_1(x)=\frac{\tau^2}{x^2\left(x^2+\frac{1}{4}\right)}\left(1-\frac{4\tau}{x^2+1}\right)+\mathcal{O}(\tau^4).
\ee
One can then perform the same steps for higher order corrections, at
each stage of the calculation keeping all the relevant terms. For
example, already at the fourth order in $\tau$ of $\varphi_1(x)$ one
cannot neglect the lowest order contribution coming from
$\varphi_2(x)$ in the r.h.s. of Eq.~(\ref{perturbative}). For higher
orders one also has to consider corrections to $\varphi_n(x)$ with
$n\geq 2$. 
\section{Small $\gamma$ limit for $g_2$}
\label{small_gamma}
In this appendix we prove that
\begin{equation}
\lim_{\gamma\to 0}g_2(\gamma)=2\ .
\label{a:to_prove}
\end{equation}
Our starting point is Eqn (\ref{new_g2}). Rescaling variables by
\begin{eqnarray}
\hat{b}_{m}(x)=\sqrt{\gamma}b_{m}\left(\frac{x}{\sqrt{\gamma}}\right)\ ,\qquad \hat{\eta}_{n}(x)=\widetilde{\eta}_{n}\left(\frac{x}{\sqrt{\gamma}}\right)\ ,
\end{eqnarray}
we have
\begin{eqnarray}
g_2=2+ \sqrt{\gamma}\sum_{m=1}^{\infty}\int_{-\infty}^{\infty}\frac{\mathrm{d}x}{2\pi}\ \left[2mx\hat{b}_m(x)\frac{1}{1+\hat{\eta}_m(x)}\right]\,.
\label{a:asympt}
\end{eqnarray}
The functions $\hat{b}_{n}(x)$ satisfy the coupled nonlinear integral equations
\begin{equation}
\hat{b}_n(x)=nx-\sum_{m=1}^{\infty}\int_{-\infty}^{\infty}\mathrm{d}y\ \frac{1}{1+\hat{\eta}_{m}(y)}\hat{b}_{m}(y)\hat{a}_{nm}(x-y)\ ,
\end{equation}
where
\begin{equation}
\hat{a}_{nm}(x)=\frac{1}{\sqrt{\gamma}}\widetilde{a}_{nm}\left(\frac{x}{\gamma}\right).
\end{equation}
Our goal is to determine the limit
\begin{equation}
\lim_{\gamma\to 0}\sum_{m=1}^{\infty}\int_{-\infty}^{\infty}\frac{\mathrm{d}x}{2\pi}\ \left[2mx\hat{b}_m(x)\frac{1}{1+\hat{\eta}_m(x)}\right]\ .
\end{equation}
The calculation is non-trivial as we cannot exchange the infinite sum 
with the  limit. However, based on numerical evidence we claim that
this limit is finite, and (\ref{a:to_prove}) then immediately follows
from (\ref{a:asympt}). 

Note that the numerical computation of $g_2(\gamma)$ is increasingly
demanding as $\gamma\to 0$, due to the fact that more and more strings
contribute. Accordingly, the infinite systems (\ref{one}) and
(\ref{two}) have to be truncated to a larger number of equations and
the numerical computation takes more time to provide precise results.  
We were able to numerically compute $g_2(\gamma)$ for decreasing
values of $\gamma$ down to $\gamma=0.025$ where $g_2(0.025)\simeq
2.11$ and approximately $30$ strings contributed to the
computation. We fitted the numerical data for small $\gamma$ with
$G(\gamma)=\alpha_1+\alpha_2\sqrt{\gamma}$ and we correctly found
$\alpha_1=2$ within the numerical error. 

\end{appendix}


\nolinenumbers

\end{document}